\documentclass[sort&compress]{elsarticle}

\usepackage[reviewcopy]{adndt}
\journal{Atomic Data Nuclear Data Tables}

\usepackage{array}
\usepackage{amssymb}
\usepackage{caption}
\usepackage{comment}
\usepackage{subcaption}
\usepackage{amsmath}    
\usepackage{graphicx}
\usepackage[colorlinks]{hyperref}
\usepackage{cleveref}
\usepackage{float}
\usepackage{longtable}
\usepackage{multirow}
\usepackage{makecell}
\usepackage{natbib}
\usepackage{threeparttable}
\usepackage{longtable}
\usepackage{tablefootnote}
\usepackage{tabulary}
\usepackage{longtable}
\usepackage{isotope}
\usepackage[version=4]{mhchem}
\usepackage{lineno}

\begin{document}

\begin{frontmatter}

\title{Revisiting the nuclear magnetic octupole moment}

\author[first]{S. Bofos\fnref{label1}}
\author[first]{T.J. Mertzimekis\corref{cor1}}
\ead{tmertzi@phys.uoa.gr}

\affiliation[first]{organization={National and Kapodistrian University of Athens},
            addressline={Zografou Campus}, 
            city={Athens},
            postcode={GR-15784}, 
            country={Greece}}

\cortext[cor1]{Corresponding author.}
\fntext[label1]{Present address: CEA DES/IRESNE/DER/SPRC/LEPh, France}

\begin{abstract}
The nuclear magnetic octupole moment is revisited as a potentially useful observable
for nuclear structure studies. The magnetic octupole moment, $\Omega$, is examined
in terms of the nuclear collective model including weak and strong coupling.
Single-particle formulation is additionally considered in the overall comparison of
theoretical predictions with available experimental data. Mirror nuclei symmetry is
examined in terms of the magnetic octupole moment isoscalar and isovector terms. A
full list of predictions for $\Omega$ of odd-proton and odd-neutron nuclei in
medium-heavy mass regimes of the nuclear chart is produced aiming at providing starting
values for future experimental endeavors.
\end{abstract}



\begin{keyword}
Nuclear magnetic octupole moment \sep
strong coupling \sep
mirror nuclei



\end{keyword}

\end{frontmatter}

\clearpage

\tableofcontents
\listofDtables
\listofDfigures

\clearpage

\section{Introduction}
\label{sec:intro}

The nucleus is conventionally described using a hierarchy of static electromagnetic moments.
The most known, magnetic dipole~\cite{2019_Stone_longlived_mu,2020_Stone_shortlived_mu} and
electric quadrupole~\cite{2021_Stone_Q} moment have been studied extensively
and have given much information about nuclear structure, while the next higher, the magnetic
octupole moment, remains relatively unexplored. A systematic study of octupole moments will
help place constraints on the poorly known isoscalar part of nuclear spin-spin
forces~\cite{Beloy2008} and in addition will reduce the uncertainty of nuclear Schiff
moments~\cite{dobaczewski2018}. The main reason for the lack of information about octupole
moments is the great difficulty in deducing the magnetic octupole moment, $\Omega$, from
measurements of the hyperfine intervals, since it requires knowing atomic-structure couplings,
that become inaccurate for multivalent atoms. There are only few cases
that special features of the nuclei enable us to extract their octupole moment from
measurements of hyperfine intervals. As a result, up to now, $\Omega$ of only 21 isotopes
have been measured~\cite{schwartz1957,amoruso1971,zacharias1955,brown1966,gerginov2009,eck1957,gerginov2003,blachman1967,landman1970,faust1963,faust1961,lewty2013,lewty2012corrected,Hoffman2014,Unsworth1969,singh2013,deGroote2021,mcdermott1960,olsmats1961,daly1954,jaccarino1954,hull1970,deGroote2022,rosen1972}.
This lack of experimental results makes the prediction of $\Omega$ extremely difficult as
there is no nuclear model that ensures the appropriate description of $\Omega$. The goal of this work is
to review the already known information about $\Omega$ and collect the available experimental
data in order to compare them with the predictions of nuclear models attempting to explore
any possible trends. This is the first attempt to compare experimental data of $\Omega$ with
the predictions of various nuclear models since the work of Suekane and Yamaguchi in
1957~\cite{Suekane1957}. In their work, the authors suggested that the strong coupling
case of the collective model is the most favorable to describe $\Omega$ properly.

\section{Definitions and models}
\label{sec:Literature Review}

The nuclear magnetic octupole moment ($\Omega$) is defined as $\Omega=-M_3$,
where $M_3$ is the expectation value of the corresponding operator~\cite{schwartz1955}
\begin{equation}
	M_{\lambda}^\mu=\mu_N\int\Psi^*\left(\mathbf{\nabla}r^{\lambda}C_{\mu}^{(\lambda)}(\theta,\phi)\right)\cdot\left(g_l\frac{2}{\lambda+1}\mathbf{L}+g_s\mathbf{S}\right)\Psi d\upsilon ,
\label{eq:general magnetic moment}
\end{equation}
where the integral is over the nuclear volume $\upsilon$ and
\[
C_\mu^{(\lambda)}(\theta,\phi)=\left(\frac{4\pi}{2\lambda+1}\right)^{1/2}Y_{\lambda\mu}(\theta,\phi) .
\]
is the multipole tensor operator of order $\lambda$ with parity $(-1)^{\lambda}$. $Y_{\lambda\mu}$ denotes the normalized spherical harmonic
function of order $\lambda$, $\mu$.

The $M_3$ operator, given in Eq.~\ref{eq:general magnetic moment}, differs in nature from the
nuclear dipole moment which is a vector. If the wave function for a nucleus with spin $I$ in
the magnetic substate $m$ is $\Psi_I^m$, then the magnetic moment of order $\lambda$ is the expectation
value of the substate with the maximum spin projection
\begin{equation}
M_{\lambda} =\langle\Psi_I^I|\hat{M}_{\lambda}|\Psi_I^I\rangle .
\end{equation}
\label{eq:Omega1}
This quantity vanishes for even $\lambda$ and $\lambda>2I$. So, it becomes clear that $\Omega\neq 0$
for values $I\geq 3/2$~\cite{Suekane1957}. A non-zero value of $\Omega$ may be related to the
octupole deformation of the nucleus. The study of the octupole deformation and its evolution
in nuclei presents great opportunities for both experimental and theoretical research,
as it can reveal important issues on symmetries of the nuclear forces (see example works
Refs.~~\cite{Zhu1995,Auerbach1996,Bonatsos2005, Minkov2006,robledo2011,Gaffney2013,Bonatsos2015,Brewer2017,Butler2020}).

The main problem with the calculation of the $\Omega$ is to find the appropriate wave
functions to describe the nucleus. Consequently, $\Omega$ can be deduced using various
models, such as:
\begin{itemize}
    \item Single--particle model
    \item Collective model
\end{itemize}
The relevant expressions for $\Omega$ based on these nuclear models are deduced below and
the corresponding predictions are compared with the experimental data, where available.
Besides these models, a mean-field approach in calculating $\Omega$ could seem a rather 
straightforward task; however, there is very limited information on such efforts in
literature~\cite{chinn1996}, stressing the need for more work along these lines.

\subsection{Single--Particle Model}
\label{sec:Shell_Model}

In the extreme single--particle model $\Omega$ is expressed as~\cite{schwartz1955}
\begin{equation}
\begin{aligned}
\Omega=&+\mu_{N}\frac{3}{2} \frac{(2 I-1)}{(2 I+4)(2 I+2)}\left\langle r^{2}\right\rangle \\
& \times\left\{\begin{array}{l}
{(I+2)\left[\left(I-3/2\right) g_{l}+g_{s}\right]} \text{ for } {I=l+\frac{1}{2}},
\end{array}\right.\\
& \times\left\{\begin{array}{l}
{(I-1)\left[\left(I+5/2\right) g_{l}-g_{s}\right]} \text{ for } {I=l-\frac{1}{2}},
\end{array}\right.
\end{aligned}
\label{eq:Omega shell model}
\end{equation}
where $g_l$ and $g_s$ are constants determined by the type of the valence nucleon. More specifically, their values are

$g_l=1$ and $g_s=+5.587$ for valence proton,

$g_l=0$ and $g_s=-3.826$ for valence neutron.

The aforementioned bare nucleon values correspond to the extreme single--particle model.
In order to account for observed deviations between experimental values and the predictions
of extreme single--particle model, effective $g$ factors that take into account configuration
mixing effects can be adopted. Typical selections of these effective values are
$g_s^{eff}=0.7g_s$ and $g_l^{eff}=g_l+\delta g_l$ where $\delta g_l=+0.1$ for unpaired proton and $\delta g_l=0.0$ for unpaired neutron. 
We will refer to this case as the {\em effective} single--particle model hereafter.

The extreme single-particle model should not be trusted for more than a rough estimation
of the value of $\Omega$, as the correct values depend strongly on nuclear many-body effects,
more specifically on core polarization mediated by the nucleon spin-spin
interaction~\cite{Beloy2008,senkov2002}. This is analogous to the Schmidt limits in
magnetic dipole moments~\cite{mertzimekis2016}.

\subsection{Collective Model}
\label{sec:Collective_Model}

The collective degrees of freedom are associated with deformations in shape with approximate
preservation of volume. The normal coordinates of these oscillations would be the expansion
parameters $\alpha_{\lambda\mu}$ of the nuclear surface defined by
\begin{equation}
    R(\theta,\phi)=R_0\left[1+\sum_{\lambda\mu}\alpha_{\lambda\mu}Y_{\lambda\mu}(\theta,\phi)\right],
    \label{eq:R(beta,gamma)}
\end{equation}
where $R_0$ is the equilibrium radius. In the basis of the collective model, the magnetic
octupole moment operator is considered comprising two separate parts:
(a) the core part, which acts on the nucleons that constitute the filled shells, and
(b) the particle part, which acts on the particles outside the core. The coupling term
of the Hamiltonian that describes the interaction of the angular momentum of the surface
with the angular momentum of the particles is given by
\begin{equation}
    H_{int}=-\sum_p{k(r_p)\sum_{\lambda\mu}\alpha_{\lambda\mu}Y_{\lambda\mu}(\theta_p,\phi_p)},
    \label{eq:H_int}
\end{equation}
where the summation over $p$ should be considered for the valence nucleons.
The term $k(r_p)$ is related to the strength of the coupling between the angular momenta.
The assumption of a sharp nuclear boundary implies that $k(r_p)$ has the form of a delta
function, allowing the use of the constant $k$~\cite{Bohr-Mottelson}. The
strong interaction between the core and the valence nucleons and the interaction of the
valence nucleons with themselves are neglected. Two extreme cases are examined in the present work, where the coupling
between the core and the particles is either strong or weak.

\subsubsection{Strong coupling case}

The form of the operator that acts on the particles is given by Eq.~\ref{eq:general magnetic moment}
with the appropriate change in the sign. The action of this operator should be summed
over all valence nucleons. As for the operator that acts on the core of the nucleus,
starting by the basic Eq.~\ref{eq:R(beta,gamma)} and considering quadrupole deformation
of the nucleus ($\lambda~=~2$) we get the expression~\cite{Suekane1957}
\begin{equation}
    \hat{\Omega}_c=-\frac{i}{28}\frac{3AM}{4\pi\hbar}g_RR_0^4\sum_{\mu=-2}^{\mu=2}{\mu(17-5\mu^2)\alpha_{2\mu}^*\Dot{\alpha}_{2\mu}},
    \label{eq:Omega_operator core}
\end{equation}
where $g_R\simeq Z/A$ is the $g$ factor of the core and $M$ is the nucleon mass.
The choice of $\lambda=2$ is not arbitrary, but based on the fact that second-order
deformations are considered more impactful on the octupole magnetic moment, in contrast with
higher order terms which are often neglected by researchers.

Using the strong coupling wave functions described in Ref.~\cite{Bohr-Mottelson} the octupole magnetic moment is obtained as
\begin{equation}
    \Omega=P_3(\Omega_{sp}+\Omega_0),
    \label{eq:Omega value strong cc}
\end{equation}
where $P_3$ and $\Omega_0$ are given by
\begin{equation}
    P_3=\frac{2I(2I-1)(2I-2)}{(2I+2)(2I+3)(2I+4)},
    \label{eq:Factor P3 for scc}
\end{equation}
\begin{equation}
    \Omega_0=-\frac{30}{49}\sqrt{\frac{2}{3}}g_RR_0^2 ,
    \label{eq:Omega0 for scc}
\end{equation}
if $j$ is assumed to be a good quantum number, with $\Omega_{sp}$ given by
Eq.~\ref{eq:Omega shell model}.

As a matter of fact, Eq.~\ref{eq:Factor P3 for scc} is correct for $I>3/2$.
When $I=3/2$ the degeneracy of two states occurs and the wave function
becomes complicated. Consequently, for $I=3/2$ the octupole moment is
expressed as
\begin{equation}
    \Omega\simeq\frac{19}{35}\Omega_{sp}+\frac{1}{35}\Omega_0
    \label{eq:Omega value strong cc for I=3/2}
\end{equation}

\subsubsection{Weak coupling case}

For the weak coupling case perturbation theory can be used, considering
$H_{int}$ as a small perturbation. Following the same procedure for the
operators as with the strong coupling case results in the expression
for the octupole moment correct up to second order in $H_{int}$ ($I=j$)
\begin{equation}
    \Omega=\Omega_p+\Omega_c ,
    \label{eq:Omega break2}
\end{equation}
where
\begin{equation}
\begin{aligned}
    \Omega_p=\Omega_{sp}
    &\left[1
    -k^2\frac{\hbar\omega}{2C}\sum_{j'}\frac{\langle j|h_2|j'\rangle^2}{(\hbar\omega+\Delta_{jj'})^2}\right.\\
    &\left.\phantom{\left[ 1\right.}+k^2\frac{\hbar\omega}{2C}\sum_{j'j''}\frac{\langle j|h_2|j''\rangle\langle j|h_2|j'\rangle}{(\hbar\omega+\Delta_{jj''})(\hbar\omega+\Delta_{jj'})}
    \times(-1)^{j'+I-3}(2I+1)W(j''Ij'I;23)
    \frac{\langle j''||\hat{\Omega}_p||j'\rangle}{\langle j||\hat{\Omega}_p||j\rangle}\right] ,
\end{aligned}
    \label{eq:Omega_p value weak cc}
\end{equation}
and
\begin{equation}
\begin{aligned}
    \Omega_c= -\frac{1}{28}R_0^2g_R\frac{k^2\hbar\omega}{C}\sum_{j'}\frac{\langle j|h_2|j'\rangle^2}{(\hbar\omega+\Delta_{jj'})^2} \times\sum_{\mu=-2}^{\mu=2}\mu(17-5\mu^2)\langle j'2I-\mu\mu|j'2II\rangle^2 .
\end{aligned}
    \label{eq:Omega_c value weak cc}
\end{equation}
Using the Bohr--Mottelson's notation,
$\Delta_{jj'}$ is the energy difference of the particle levels with $j$ and $j'$,
$k$ is the coupling constant as defined in Eq.~\ref{eq:H_int},
$W$ are the Racah coefficients,
$\langle j|h_2|j'\rangle=\langle j||Y_2||j'\rangle/\sqrt{2j+1}$,
$\Omega_{sp}$ is given by Eq. \ref{eq:Omega shell model}
and $C$ is the nuclear deformability~\cite{Bohr-Mottelson}
\begin{equation}
    \label{eq:deformability constant C}
    C_{\lambda}=(\lambda-1)(\lambda+2)R_0^2S-\frac{3}{2\pi}\frac{\lambda-1}{2\lambda+1}\frac{Z^2e^2}{R_0} ,
\end{equation}
where $R_0$ is the radius of the nucleus and $S$ is the surface tension.
The analysis of nuclear binding energies leads to the estimate
$4\pi R_0^2S=15.4~A^{2/3}$ MeV~\cite{rosenfeld1948} and
$\omega$ is the oscillation frequency of the surface
\begin{equation}
    \label{eq:omega frequency of surface in CM}
    \omega=\sqrt{\frac{C_{\lambda}}{B_{\lambda}}} ,
\end{equation}
where using the classical hydrodynamical expression
$B_{\lambda}=\frac{1}{\lambda}\frac{3}{4\pi}AMR_0^2$~\cite{Bohr-Mottelson}.
When $\Delta_{jj'}(j'\neq j)$ is large and $j$ may be considered a good quantum number,
$\Omega_{c}$ and $\Omega_{p}$ are simplified as
\begin{equation}
    \Omega_{c}=-\frac{45}{224 \pi} g_{R} R_{0}^{2} \frac{k^{2}}{\hbar \omega C} \cdot \frac{(2 I-1)(I-1)}{I(I+1)^{2}},
    \label{eq:Omega_c value weak cc simplified version}
\end{equation}
and
\begin{equation}
    \Omega_{p}=\Omega_{sp}\left[1+\frac{5}{32 \pi} \frac{k^2}{\hbar \omega C}
    \left(\mathcal{F}_1-\mathcal{F}_2\right)\right],
    \label{eq:Omega_p value weak cc simplified version}
\end{equation}
where
\begin{equation}
    \mathcal{F}_1 = \frac{4 I^2(I+1)^2-75 I(I+1)+234}{4 I^2(I+1)^2},
\end{equation}
\begin{equation}
    \mathcal{F}_2 = \frac{(2 I-1)(2 I+3)}{4 I(I+1)}.
\end{equation}

\section{Comparison with experimental results and predicted values for unmeasured cases}
\label{sec:comp}

\begin{figure*}[ht]
    \centering
    \includegraphics[width=0.98\textwidth]{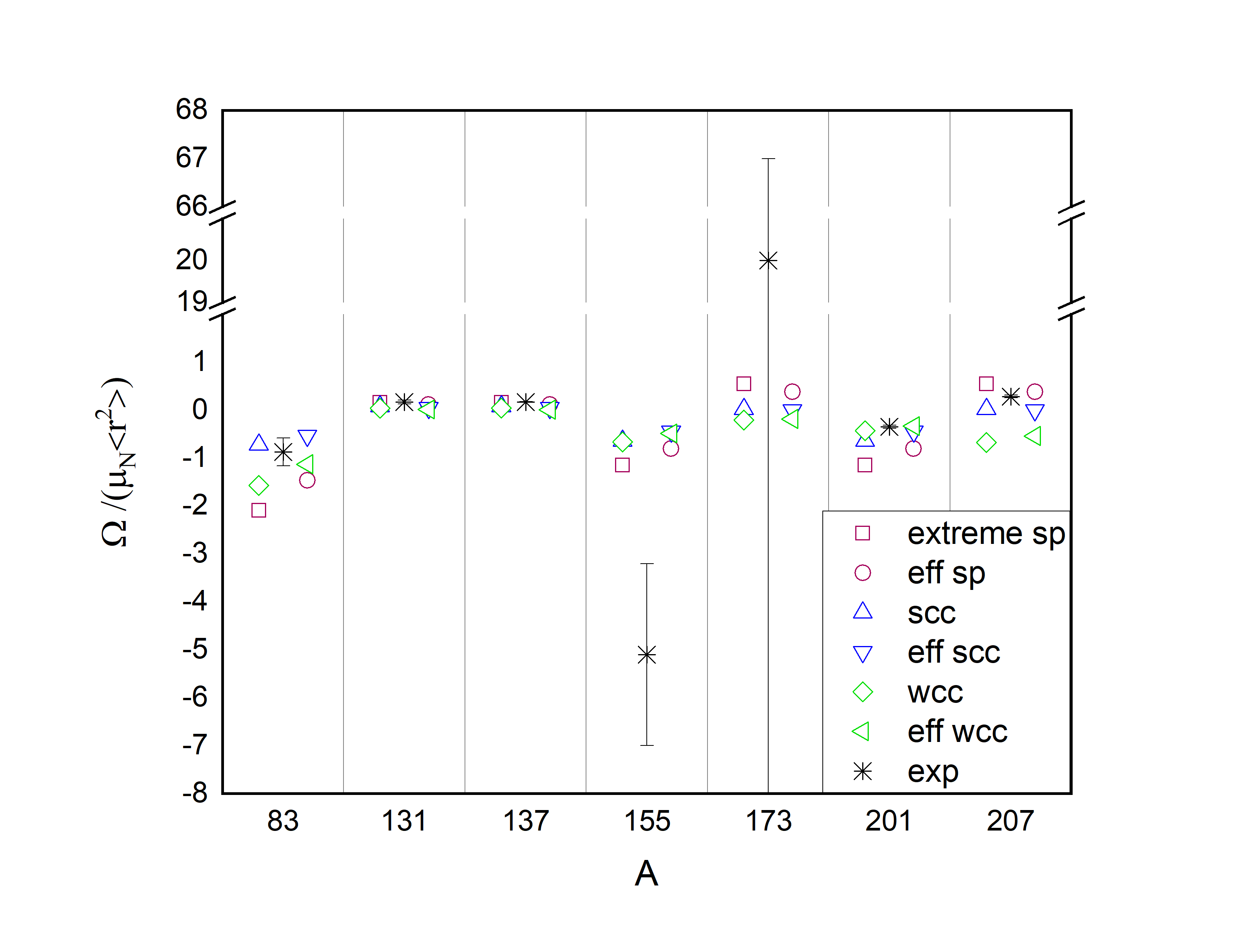}
    \caption{Visual comparison of experimental results ({\em exp}) for $\Omega$ and the
    predictions of extreme and effective ({\em eff}) single particle model
    ({\em sp}, magenta), strong ({\em scc}, blue) and weak ({\em wcc}, green) coupling case
    for nuclei with unpaired neutron, respectively. For better visualization, the points of
    {\em eff sp}, {\em eff scc} and {\em eff} {\em wcc} have been moved slightly to the right.
    The case of \ce{^{173}Yb} is not presented in
    the diagram. For some isotopes the error bars are smaller than the size of the plotted points.
    Please note that the x-axis is not linear, while the y-axis includes breaks at two values to include the distinctively high value of \ce{^{173}Yb} together with the rest of the plotted data.}
    \label{fig:Om_Sum_up_neutron}
\end{figure*}

\begin{figure*}[ht]
    \centering
    \includegraphics[width=0.98\textwidth]{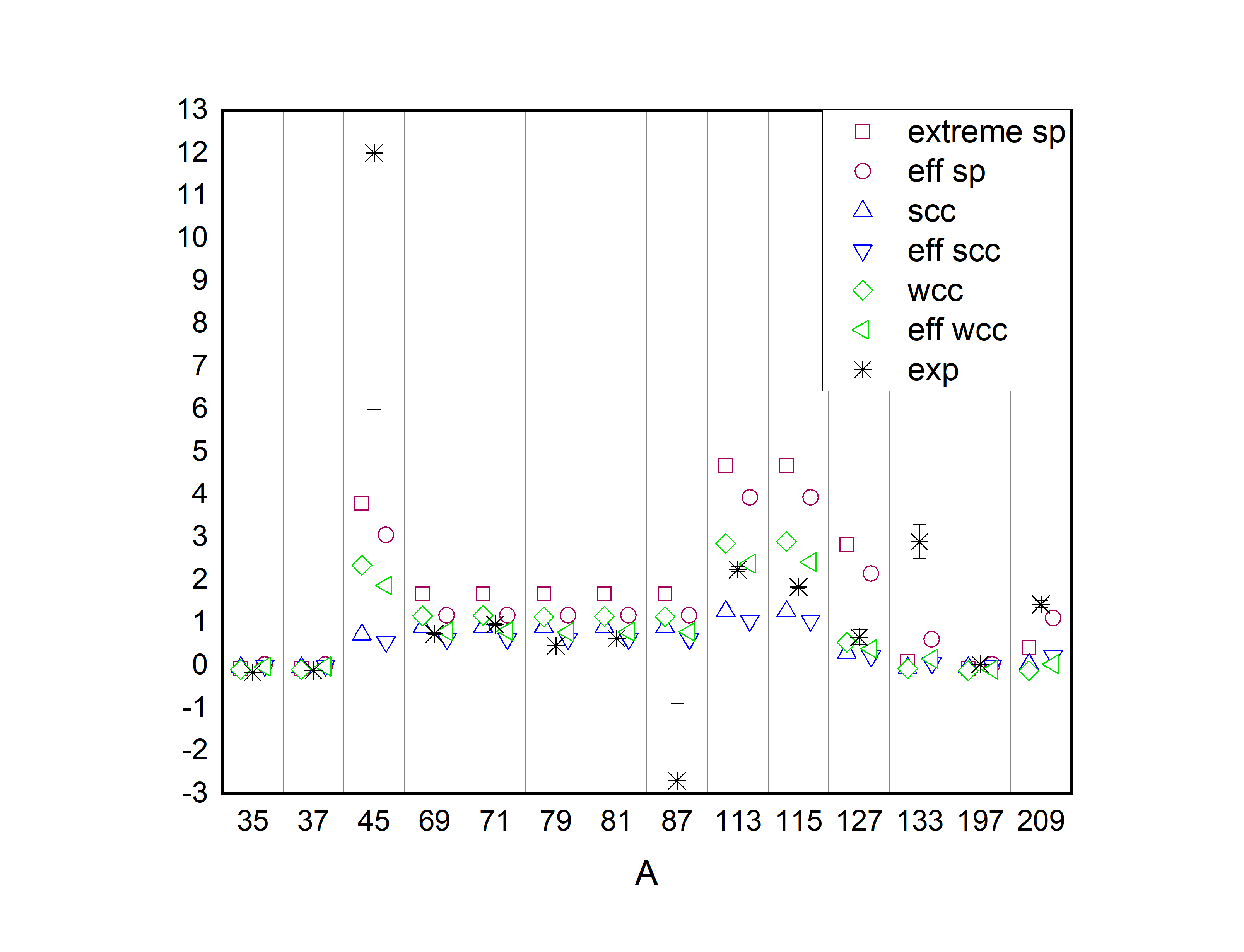}
    \caption{Visual comparison of experimental results ($exp$) for $\Omega$ and the predictions of
    extreme and effective ({\em eff}) single particle model ({\em sp}, magenta),
    strong ({\em scc}, blue) and
    weak ({\em wcc}, green) coupling case
    for nuclei with an unpaired proton, respectively. For better visualization, the points of
    {\em eff sp}, {\em eff scc} and {\em eff} {\em wcc} have been moved slightly to the right.
    For some isotopes the error bars are smaller
    than the size of the plotted points. The values for \ce{^{79}Br} and \ce{^{81}Br} have no cited
    uncertainty in literature. Please note that the x-axis is not linear.}
    \label{fig:Om_Sum_up_proton}
\end{figure*}

After this brief presentation of the models used for the deduction of $\Omega$ a comparison is displayed between the predictions of each model and the very limited experimental results that
are currently available. Experimental data have been retrieved from Fuller's 1976
compilation~\cite{fuller1976} and cross-checked with the values published in the original
works. Those data, as well as additional recent measurements that were collected from
literature are about to be publicly available soon in the nuclear moments database,
NUMOR~\cite{mertzimekis2016NUMOR,2024_Mertzimekis}. This assemble, to our knowledge,
is the most complete collection of $\Omega$ measurements. The predictions of each model
along with the experimental values for the isotopes that their octupole moment has already
been measured are presented in Tables~\ref{tab:valence_neutron} and \ref{tab:valence_proton}
and in Figs.~\ref{fig:Om_Sum_up_neutron} and \ref{fig:Om_Sum_up_proton}, as well. Some of
the published data do not quote an experimental uncertainty and are marked with a
$b$ in Table~\ref{tab:valence_proton}.

We have used $\langle r^2\rangle=3/5R_0^2$, $R_0=1.35A^{1/3}$~fm~\cite{seth1958},
and $k=40$~MeV, while $C$, $\hbar\omega$ have been calculated from
Eqs.~\ref{eq:deformability constant C} and \ref{eq:omega frequency of surface in CM}.

As seen in Tables~\ref{tab:valence_neutron} and \ref{tab:valence_proton} the strong coupling
case predicts the correct sign for most of the isotopes (except \ce{^{133}Cs} 
and \ce{^{197}Au}). The same is true for the effective strong coupling case with the only exception being \ce{^{37}Cl}. The case of \ce{^{87}Rb} is special because many--body perturbation theory
that is used for the extraction of $\Omega$ from the hyperfine interaction constant $c$ results 
in the different sign between the experimental value and the predictions of each
model~\cite{gerginov2009}.

In a similar way, the weak coupling case predicts the correct sign for most of the isotopes
(except \ce{^{133}Cs}, \ce{^{197}Au}, \ce{^{207}Po} and \ce{^{209}Bi}). The same is true for
the effective weak coupling case with the only exceptions being \ce{^{197}Au} and \ce{^{207}Po}.
Moreover the strong coupling and the effective weak coupling case predicted exceptionally well
the order of magnitude of $\Omega$ for isotopes whose unpaired nucleon has spin $I=l+1/2$, except
\ce{^{45}Sc} and \ce{^{155}Gd}. However, in the case of \ce{^{155}Gd}, the experimental value
is only considered tentative~\cite{Unsworth1969}. The failure in describing isotopes whose
unpaired nucleon has spin $I=l-1/2$ could be due to a cancellation between the orbital and spin
contributions of the valence nucleon  to the magnetic octupole moment in the basis of single
particle model~\cite{Beloy2008}. This observation, combined with the comment by Schwartz~\cite{schwartz1957}
that until more is known about polarization effects, the experimental values cannot be trusted
for more than a qualitative comparison with the nuclear models, makes the strong coupling and
the effective weak coupling cases very promising models for the description of $\Omega$ for
nuclei with spin $I=l+1/2$.

It might be useful to examine a different approach in the grouping of the isotopes, based on which model or models should be applicable. One way to classify the nuclei is based on the $R_{4/2}$
ratio in the even-even neighbor of the nuclide of interest. If $R_{4/2}<1$, then a single-particle model
should be most applicable; for $R_{4/2}\sim2$, weak coupling is likely to be most appropriate.
The strong coupling case should become applicable when $R_{4/2}\sim10/3$~\cite{casten1993}.
Despite the fact that the distinction is not always clear cut the isotopes were divided into these three groups.

The first group is consisted by the isotopes \ce{^{87}Rb} and \ce{^{209}Bi} for which a single-particle model
is expected to be most applicable. Based on the experimental values this is true, as either the extreme
or the effective case of the single-particle model is the most accurate in their prediction.

The second group is consisted by isotopes \ce{^{155}Gd} and \ce{^{173}Yb} with $R_{4/2}>3$.
This value indicates well-deformed nucleus, consequently the strong coupling case is expected to be the most
appropriate. While for \ce{^{155}Gd} there is no success by any model, the case of \ce{^{173}Yb} is different.
In a recent paper published by de Groote {\em et al.}~\cite{deGroote2021} there is no significant evidence for a non-zero octupole moment in \ce{^{173}Yb}. As a result,
the effective strong coupling case seems accurate in its prediction.

The third group is consisted by the remaining isotopes mentioned in Tables~\ref{tab:valence_neutron} and
\ref{tab:valence_proton}. For these isotopes the weak coupling case is expected to be the most applicable.
While this is true for the majority of odd-proton nuclei, for odd-neutron nuclei the expected model based
on $R_{4/2}$ ratio has only occasionally predicted the correct order of magnitude for $\Omega$. More
specifically, for nuclei \ce{^{131}Xe}, \ce{^{137}Ba}$^+$ and \ce{^{207}Po} extreme or effective
single-particle model predicted accurately their $\Omega$ value in contrast with the weak coupling case.
The same situation is observed for \ce{^{197}Au}, while for nuclei \ce{^{45}Sc} and \ce{^{133}Cs} neither
the weak coupling case nor any other model had any success in their prediction.

It is worth noting that both strong and weak coupling cases are successful in predicting
the order of magnitude in several isotopes, with the weak coupling case having greater
deviation from the experimental values. Interestingly, the opposite is true when the effective $g$
factors have been incorporated in the calculations. The effective weak coupling case is clearly
more successful in describing experimental data with respect to the effective strong coupling case
(7 vs 2 isotopes). The prediction of the order of magnitude for the effective strong and weak
coupling cases for isotopes with unpaired proton in subshell $2p_{3/2}$, \ce{^{69,71}Ga}
and \ce{^{79,81}Br} --except \ce{^{87}Rb}, was also very good. In addition, the effective weak
and strong coupling cases correctly predicted the order of magnitude for \ce{^{83}Kr}, \ce{^{127}I}
and \ce{^{201}Hg}. These isotopes have more than 3 valence nucleons so the success could be possibly
attributed to the increased impact of collective phenomena when moving away from magic numbers.
However, there are other isotopes with 3 or more valence nucleons, such as \ce{^{35,37}Cl} and
\ce{^{197}Au} that are not described correctly by the collective model.

Finally, there is a discrepancy of an order of magnitude between our theoretical predictions
for $\Omega_c$ (Eq. \ref{eq:Omega_c value weak cc simplified version}) and those originally
reported by Suekane and Yamaguchi~\cite{Suekane1957}, for isotopes \ce{^{115}In} and \ce{^{127}I}.

The effective single particle model predictions for the majority
of isotopes are more accurate than those for the extreme model. Moreover, the effective model
predicted the correct order of magnitude of $\Omega$ for most of the isotopes.
Another interesting observation is that the effective single particle model predicted successfully
the order of magnitude of $\Omega$ for all the isotopes with $A>196$. All of them have 3 or less
valence nucleons, so the success of the effective single particle model could be due to their
proximity to a closed-shell configuration. Most of these isotopes have spin $I=l-1/2$ and the
effective single particle is the only model that has some accuracy in their prediction.

There is a general trend of the values of $\Omega$ to fall inside the Schmidt lines of extreme
single-particle model. However, there are isotopes such as \ce{^{45}Sc}, \ce{^{155}Gd} and \ce{^{173}Yb}
that do not follow this trend and stress the need for further study, at both theoretical and
experimental level.

It should be noted that both \ce{^{155}Gd} and \ce{^{173}Yb} are well-deformed nuclei in regions
where their even-even cores are strongly collective with $R_{4/2}>3$. In contrast, \ce{^{45}Sc}
nucleus with a single proton outside of the magic shell of $Z=20$ should be near spherical and
calculable with the shell model. In \cite{deGroote2022} it is clear that despite the structure
of \ce{^{45}Sc} shell model calculations, including different interactions in a $(sd)pf$-shell
model space, cannot reproduce its experimental value of $\Omega$.

In general, the usually smaller absolute value of experimental results compared to the extreme
single-particle model predictions may be explained either by the configuration mixing method or
the collective model.

If one takes the former, one has to introduce new configurations into the ground state which
contribute to $\Omega$, but not to $\mu$ and $Q$~\cite{Suekane1957}. In a paper by Brown
{\em et al.}~\cite{brown1980}, $\Omega$ of isotopes \ce{^{35}Cl} and \ce{^{37}Cl} have been calculated
from a complete $sd$-shell space wave functions using three sets of $g$ factors. Their predictions
using the free-nucleon $g$ factors are in better agreement with the experimental values.

From the comparison of extreme single particle model and single particle model
using effective $g$ factors with the available experimental data it seems that
the latter case describes the experimental values more accurately.
Again, configuration mixing effects~\cite{noya1958,van1981,wolf1987,vergados1971,mertzimekis2003,stuchbery1991}
can be possibly the reason.
At this point it should be mentioned that we used effective $g$ factors that are
typically used in the $M_1$ operator. The effective $g$ factors in the $M_3$ operator
need not be identical to those in the $M_1$ operator, but they will be related to
the extent that they describe the same configuration mixing effects.
It is reasonable to start by assuming that they are the same.

For the strong coupling case the reduction of $\Omega$ is achieved via the projection
factor $P_3$ as seen in Eq.~\ref{eq:Omega value strong cc}, which represents the
transformation to the space fixed frame from the body fixed frame. The extremely
limited experimental information available for $\Omega$ available in 1957 were fitted
by the collective model, using the strong coupling case. The additional experimental
results that are available today seem to give the edge to the effective weak
coupling case.

Concluding the application of the formalism discussed above, a complete set of
values of the magnetic octupole moment for valence protons in subshells $2p_{3/2}$
and $1g_{9/2}$, as well as valence neutrons in $1g_{9/2}$ has been produced. The
values are shown in detail in Tables~\ref{tab:p2p3_2}, \ref{tab:p1g9_2} and
\ref{tab:n1g9_2}
where we have used the bare nucleon $g$ factors, exclusively.

\section{Octupole moments and mirror nuclei}
\label{sec:mirror}

It is known that the sum of the ground-state magnetic moments of mirror nuclei can
give valuable information about gross-spin properties of nuclei. More specifically,
systematic trends in the spin expectation value as a function of the mass number
for all $T=1/2$, $3/2$ and $5/2$
nuclei~\cite{buck1983,buck2001,hanna1985,mertzimekis2006,berryman2009,perez2008,minamisono2006,mertzimekis2016}
provide us with a valuable tool in predicting ground-state dipole magnetic moments
of mirror partners. In a similar framework, a systematic study of magnetic octupole
moments of mirror nuclei could reveal similar trends that have the potential to
serve as a quick way in predicting $\Omega$ of a nucleus if its mirror partner value
is known experimentally.

The magnetic moment of a nucleus can be expressed in a formal way in the framework
of the one-body magnetic moment operator with the help of the isospin formalism. By
extending the work of Sugimoto~\cite{sugimoto1969,mertzimekis2016} we will express
$\Omega$ as an isoscalar and an isovector term.

Using the standard notation about expectation values, $\Omega$ can be expressed as
\begin{equation}
\begin{aligned}
    \Omega=\left\langle\sum_{i}\left[\frac{1}{2}\left(1+\tau_{3}^{i}\right)\left(\mathbf{\nabla}r^3C_{3}^{(3)}(\theta,\phi)\right)_{z} \left(\frac{l_{z}^{i}}{2}+\sigma_{z}^{i} \mu_{p}\right)
    +\frac{1}{2}\left(1-\tau_{3}^{i}\right)
    \left(\mathbf{\nabla}r^3C_{3}^{(3)}(\theta,\phi)\right)_{z} \sigma_{z}^{i} \mu_{n}\right]\right\rangle_{M=I} .
\end{aligned}
    \label{eq:Omega in isospin framework}
\end{equation}
where $\sigma$ stands for Pauli matrices, $\tau$ for isospin (and its projection
$\tau_{3}$) and $\mu_{p}$ and $\mu_{n}$ denote the dipole magnetic moment of a free
proton and neutron, respectively. The summation in $i$ should be taken over all
nucleons. This formula can be rewritten with an isoscalar and an isovector term
\begin{equation}
    \Omega=\left\langle\sum_{i}\Omega_0^{(i)}\right\rangle_{M=I}
    +\left\langle\sum_{i}\Omega_3^{(i)}\right\rangle_{M=I} ,
    \label{eq:Omega to isoscalar/isovector}
\end{equation}
where
\begin{equation}
    \Omega_0^{(i)}:=\frac{1}{2}\left(\mathbf{\nabla}r^3C_{3}^{(3)}(\theta,\phi)\right)_{z}\left[\frac{l_{z}^{i}}{2}+\left(\mu_p+\mu_n\right)\sigma_z^{(i)}\right],
    \label{eq:Omega isoscalar}
\end{equation}
\begin{equation}
    \Omega_3^{(i)}:=\frac{1}{2}\left(\mathbf{\nabla}r^3C_{3}^{(3)}(\theta,\phi)\right)_{z}\left[\tau_{3}^{i}\frac{l_{z}^{i}}{2}+\left(\mu_p-\mu_n\right)\tau_{3}^{i}\sigma_z^{(i)}\right].
    \label{eq:Omega isovector}
\end{equation}
Under the assumptions that the isospin is a good quantum number, there is charge
symmetry in nuclear forces, and the Coulomb interaction can be neglected, for a
pair of mirror states the expectation values $\left\langle\sum_{i}\Omega_0^{(i)}\right\rangle_{M=I}$
are independent of $T_3$, while the expectation values 
$\left\langle\sum_{i}\Omega_3^{(i)}\right\rangle_{M=I}$
differ only in sign for $T_3=\pm T$. Therefore, the sum of octupole moments of
mirror states
\begin{equation}
\begin{aligned}
    \label{eq:Sum of Om_for mirror nuclei}
   \Omega(T_3=+T)+\Omega(T_3=-T)
    &=2\left\langle T,T_3=+T\left|\sum_{i}\Omega_0^{(i)}\right| T,T_3=+T\right\rangle ,\\
    &=\frac{1}{2}\left\langle\sum_{i} \left(\mathbf{\nabla}r^3C_{3}^{(3)}(\theta,\phi)\right)_{z}l_z^{(i)}\right\rangle_{M=I} +(\mu_p+\mu_n)\left\langle\sum_{i}\left(\mathbf{\nabla}r^3C_{3}^{(3)}(\theta,\phi)\right)_{z}\sigma_z^{(i)}\right\rangle_{M=I} .
\end{aligned}
\end{equation}
The operators $(\mathbf{\nabla}r^3C_{3}^{(3)}(\theta,\phi))_{z}$ and $\sigma_z^{(i)}$
are acting in different Hilbert spaces, therefore their expectation
values can be separated. The goal is to express the sum of $\Omega$ of mirror states as a function
of spin expectation value $\langle\sum_{i}\sigma_z^{(i)}\rangle_{M=I}$ or differently
written $\left\langle\sigma\right\rangle$. In order to achieve that, it is essential to
express the expectation value of the operator
$\sum_i\left(\mathbf{\nabla}r^3C_{3}^{(3)}(\theta,\phi)\right)_{z}l_z^{(i)}$
as a function of $\left\langle\sigma\right\rangle$ and
$\left\langle\left(\mathbf{\nabla}r^3C_{3}^{(3)}(\theta,\phi)\right)_{z}\right\rangle$.

This may be solved by using the Racah technique with the use of which the Eq.~54 in
Ref.~\cite{schwartz1955} is derived for the matrix elements
\[
\left\langle l\frac{1}{2}I \bigg| \mathbf{\nabla}\left(gC_k\right)\right.\cdot \left. \left(g_l\mathbf{L}+g_s\mathbf{S}\right) \bigg| l'\frac{1}{2}I'\right\rangle ,
\]
where $g$ is any function of $r$.

By defining $\Xi$ as the expectation value of the matrix that is composed of these
matrix elements, we can derive the following relations
\begin{equation}
\begin{aligned}
    \label{eq:break of expectation values}
   & \left\langle\quad\bigg|\left(\mathbf{\nabla}r^3C_{3}^{(3)}(\theta,\phi)\right)_{z}g_ll_z^{(i)}\bigg|\quad\right\rangle
   = \Xi- \left\langle\quad\bigg|\left(\mathbf{\nabla}r^3C_{3}^{(3)}(\theta,\phi)\right)_{z}\bigg|\quad\right\rangle \frac{g_s}{2}\left\langle\sigma\right\rangle .
\end{aligned}
\end{equation}
It is easy to understand that there will be a linear relation between $\langle\sigma\rangle$
and the sum of $\Omega$ expressed in Eq.~\ref{eq:Sum of Om_for mirror nuclei}.
Consequently, knowing the exact values of magnetic octupole moment for pairs
of mirror nuclei could reveal systematic trends in the spin expectation value
as a function of the mass number for nuclei with the same isospin. As a result,
the cases of \ce{^{35}Ar} and \ce{^{37}Ca} are particularly interesting for measuring,
as they are the mirror nuclei of the already measured \ce{^{35,37}Cl}. As of now,
there is no experimental information about $\Omega$ of any pair of mirror nuclei
that could ascertain this conversation. Still, the discussion is considered useful
for future experimental and theoretical research along these lines.

\section{Conclusions}
\label{sec:concl}

The octupole moment is revisited as a nuclear observable which can draw more attention
in the near future at both experimental and theoretical level. Old and recent
experimental data of magnetic octupole moments have been searched and collected
in an updated compilation.

Theoretical modeling seems to suggest that the collective model is more favorable
to describe the presently available experimental $\Omega$ data compared to the extreme
single-particle model. Both strong and weak coupling cases predict the correct sign of
$\Omega$ for the vast majority of nuclei. More importantly, both of these cases predict
correctly the order of magnitude of $\Omega$ for the majority of nuclei with spin
$I=l+1/2$. In particular, nuclei with valence proton in the $2p_{3/2}$ or in the
$1g_{9/2}$ subshell seem to be described slightly better by considering the effective
weak coupling case. The same is true for \ce{^{127}I} with valence protons in subshell
$2d_{5/2}$. More data in this subshell are required to attempt a more general
conclusion.

Furthermore, the case of \ce{^{45}Sc}, a nucleus with valence proton in subshell
$1f_{7/2}$ is not described correctly by either the strong or the weak coupling
cases. This is in agreement with recent investigations by de Groote
{\em et al.}~\cite{deGroote2022}. Finally, the strong coupling scheme seems to
deviate more than the effective weak coupling one for about 60\% of the nuclei
with spin $I=l+1/2$. As a result, despite it is quite difficult to give the edge
to either of them, using the most recent and updated measurements, the effective weak coupling case seems to gain ground in the theoretical description
over the strong coupling case, challenging the conclusion drawn by Suekane and
Yamaguchi~\cite{Suekane1957}.

At this point it is important to highlight that the majority of the isotopes, whose $\Omega$
has been measured, are in regions where their even-even cores are weakly collective with
$R_{4/2}\sim2$. One should expect that effective weak coupling case would be the most
successful model in describing these isotopes. While, this is true for odd-proton nuclei,
for odd-neutron nuclei the expected model based on $R_{4/2}$ ratio has only occasionally
predicted the correct order of magnitude for $\Omega$.

This work has additionally produced a set of predictions for more than 100 nuclei,
which are proposed as starting values for future experimental studies. As the
relevant experimental methods in radioactive beam facilities gain more power and
efficiency in accessing such nuclear observables, this list may be useful for
researchers in nuclear structure, both experimentally and theoretically.

\section*{Acknowledgements}

\noindent We thank Dr. D. Bonatsos for providing feedback on the manuscript and V.~Bofos for useful discussions.

\bibliographystyle{elsarticle-num-names} 
\bibliography{biblio_octupole}

\begin{thebibliography}{66}
\expandafter\ifx\csname natexlab\endcsname\relax\def\natexlab#1{#1}\fi
\providecommand{\url}[1]{\texttt{#1}}
\providecommand{\href}[2]{#2}
\providecommand{\path}[1]{#1}
\providecommand{\DOIprefix}{doi:}
\providecommand{\ArXivprefix}{arXiv:}
\providecommand{\URLprefix}{URL: }
\providecommand{\Pubmedprefix}{pmid:}
\providecommand{\doi}[1]{\href{http://dx.doi.org/#1}{\path{#1}}}
\providecommand{\Pubmed}[1]{\href{pmid:#1}{\path{#1}}}
\providecommand{\bibinfo}[2]{#2}
\ifx\xfnm\relax \def\xfnm[#1]{\unskip,\space#1}\fi
\bibitem[{Stone(2019)}]{2019_Stone_longlived_mu}
\bibinfo{author}{N.~J. Stone}, \bibinfo{title}{Table of Recommended Nuclear
  Magnetic Dipole Moments: Part I - Long-lived States},
  \bibinfo{type}{Technical Report}, IAEA, \bibinfo{year}{2019}.
  \DOIprefix\doi{0.61092/iaea.yjpc-cns6}.
\bibitem[{Stone(2020)}]{2020_Stone_shortlived_mu}
\bibinfo{author}{N.~J. Stone}, \bibinfo{title}{Table of Recommended Nuclear
  Magnetic Dipole Moments - Part II, Short-lived States},
  \bibinfo{type}{Technical Report}, IAEA, \bibinfo{year}{2020}.
  \DOIprefix\doi{10.61092/iaea.1p48-p6c6}.
\bibitem[{Stone(2021)}]{2021_Stone_Q}
\bibinfo{author}{N.~J. Stone}, \bibinfo{title}{Table of Nuclear Electric
  Quadrupole Moments}, \bibinfo{type}{Technical Report}, IAEA,
  \bibinfo{year}{2021}. \DOIprefix\doi{10.61092/iaea.a6te-dg7q}.
\bibitem[{Beloy et~al.(2008)Beloy, Derevianko, Dzuba, Howell, Blinov, and
  Fortson}]{Beloy2008}
\bibinfo{author}{K.~Beloy}, \bibinfo{author}{A.~Derevianko},
  \bibinfo{author}{V.~A. Dzuba}, \bibinfo{author}{G.~T. Howell},
  \bibinfo{author}{B.~B. Blinov}, \bibinfo{author}{E.~N. Fortson},
\newblock \bibinfo{title}{Nuclear magnetic octupole moment and the hyperfine
  structure of the $5{D}_{3/2,5/2}$ states of the {Ba}$^+$ ion},
\newblock \bibinfo{journal}{Phys. Rev. A} \bibinfo{volume}{77}
  (\bibinfo{year}{2008}) \bibinfo{pages}{052503}.
  \DOIprefix\doi{10.1103/PhysRevA.77.052503}.
\bibitem[{Dobaczewski et~al.(2018)Dobaczewski, Engel, Kortelainen, and
  Becker}]{dobaczewski2018}
\bibinfo{author}{J.~Dobaczewski}, \bibinfo{author}{J.~Engel},
  \bibinfo{author}{M.~Kortelainen}, \bibinfo{author}{P.~Becker},
\newblock \bibinfo{title}{Correlating {Schiff} moments in the light actinides
  with octupole moments},
\newblock \bibinfo{journal}{Phys. Rev. Lett.} \bibinfo{volume}{121}
  (\bibinfo{year}{2018}) \bibinfo{pages}{232501}.
  \DOIprefix\doi{10.1103/PhysRevLett.121.232501}.
\bibitem[{Schwartz(1957)}]{schwartz1957}
\bibinfo{author}{C.~Schwartz},
\newblock \bibinfo{title}{Theory of hyperfine structure},
\newblock \bibinfo{journal}{Phys. Rev.} \bibinfo{volume}{105}
  (\bibinfo{year}{1957}) \bibinfo{pages}{173--183}.
  \DOIprefix\doi{10.1103/PhysRev.105.173}.
\bibitem[{Amoruso and Johnson(1971)}]{amoruso1971}
\bibinfo{author}{M.~J. Amoruso}, \bibinfo{author}{W.~R. Johnson},
\newblock \bibinfo{title}{Relativistic one-electron calculations of shielded
  atomic hyperfine constants},
\newblock \bibinfo{journal}{Phys. Rev. A} \bibinfo{volume}{3}
  (\bibinfo{year}{1971}) \bibinfo{pages}{6--12}.
  \DOIprefix\doi{10.1103/PhysRevA.3.6}.
\bibitem[{Zacharias et~al.(1955)Zacharias, King, Searle, Ketudat, Vessot, Babb,
  Bates, Brown~Jr, DiBartolo, Edmonds~Jr et~al.}]{zacharias1955}
\bibinfo{author}{J.~R. Zacharias}, \bibinfo{author}{J.~G. King},
  \bibinfo{author}{C.~L. Searle}, \bibinfo{author}{S.~Ketudat},
  \bibinfo{author}{R.~F.~C. Vessot}, \bibinfo{author}{D.~D. Babb},
  \bibinfo{author}{V.~J. Bates}, \bibinfo{author}{H.~H. Brown~Jr},
  \bibinfo{author}{B.~DiBartolo}, \bibinfo{author}{D.~S. Edmonds~Jr}, et~al.,
\newblock \bibinfo{title}{Vii. atomic beams},
\newblock \bibinfo{journal}{Quarterly Progress Report}  (\bibinfo{year}{1955}).
\bibitem[{Brown and King(1966)}]{brown1966}
\bibinfo{author}{H.~H. Brown}, \bibinfo{author}{J.~G. King},
\newblock \bibinfo{title}{Hyperfine structure and octopole interaction in
  stable bromine isotopes},
\newblock \bibinfo{journal}{Phys. Rev.} \bibinfo{volume}{142}
  (\bibinfo{year}{1966}) \bibinfo{pages}{53--59}.
  \DOIprefix\doi{10.1103/PhysRev.142.53}.
\bibitem[{Gerginov et~al.(2009)Gerginov, Tanner, and Johnson}]{gerginov2009}
\bibinfo{author}{V.~Gerginov}, \bibinfo{author}{C.~E. Tanner},
  \bibinfo{author}{W.~R. Johnson},
\newblock \bibinfo{title}{Observation of the nuclear magnetic octupole moment
  of $^{87}${Rb} from spectroscopic measurements of hyperfine intervals},
\newblock \bibinfo{journal}{Can. J. Phys.} \bibinfo{volume}{87}
  (\bibinfo{year}{2009}) \bibinfo{pages}{101--104}.
  \DOIprefix\doi{10.1139/p08-145}.
\bibitem[{Eck and Kusch(1957)}]{eck1957}
\bibinfo{author}{T.~G. Eck}, \bibinfo{author}{P.~Kusch},
\newblock \bibinfo{title}{Hfs of the $5^{2}p_{\frac{3}{2}}$ state of
  {In}$^{115}$ and {In}$^{113}$: Octupole interactions in the stable isotopes
  of indium},
\newblock \bibinfo{journal}{Phys. Rev.} \bibinfo{volume}{106}
  (\bibinfo{year}{1957}) \bibinfo{pages}{958--964}.
  \DOIprefix\doi{10.1103/PhysRev.106.958}.
\bibitem[{Gerginov et~al.(2003)Gerginov, Derevianko, and Tanner}]{gerginov2003}
\bibinfo{author}{V.~Gerginov}, \bibinfo{author}{A.~Derevianko},
  \bibinfo{author}{C.~Tanner},
\newblock \bibinfo{title}{Observation of the nuclear magnetic octupole moment
  of {Cs}-133},
\newblock \bibinfo{journal}{Phys. Rev. Lett.} \bibinfo{volume}{91}
  (\bibinfo{year}{2003}) \bibinfo{pages}{072501}.
  \DOIprefix\doi{10.1103/PhysRevLett.91.072501}.
\bibitem[{Blachman et~al.(1967)Blachman, Landman, and Lurio}]{blachman1967}
\bibinfo{author}{A.~G. Blachman}, \bibinfo{author}{D.~A. Landman},
  \bibinfo{author}{A.~Lurio},
\newblock \bibinfo{title}{Hyperfine structure and ${g}_{J}$ value of the
  $^{2}{D}_{\frac{3}{2}}$ and of the $^{4}{F}_{\frac{9}{2}}$ states of
  {Au}$^{197}$},
\newblock \bibinfo{journal}{Phys. Rev.} \bibinfo{volume}{161}
  (\bibinfo{year}{1967}) \bibinfo{pages}{60--67}.
  \DOIprefix\doi{10.1103/PhysRev.161.60}.
\bibitem[{Landman and Lurio(1970)}]{landman1970}
\bibinfo{author}{D.~A. Landman}, \bibinfo{author}{A.~Lurio},
\newblock \bibinfo{title}{Hyperfine structure of the ${(6p)}^{3}$ configuration
  of {Bi}$^{209}$},
\newblock \bibinfo{journal}{Phys. Rev. A} \bibinfo{volume}{1}
  (\bibinfo{year}{1970}) \bibinfo{pages}{1330--1338}.
  \DOIprefix\doi{10.1103/PhysRevA.1.1330}.
\bibitem[{Faust and Chow~Chiu(1963)}]{faust1963}
\bibinfo{author}{W.~L. Faust}, \bibinfo{author}{L.~Y. Chow~Chiu},
\newblock \bibinfo{title}{Hyperfine structure of the metastable
  ${(4p)}^{5}(5s)^{3}{P}_{2}$ state of $_{36}\mathrm{Kr}^{83}$},
\newblock \bibinfo{journal}{Phys. Rev.} \bibinfo{volume}{129}
  (\bibinfo{year}{1963}) \bibinfo{pages}{1214--1220}.
  \DOIprefix\doi{10.1103/PhysRev.129.1214}.
\bibitem[{Faust and McDermott(1961)}]{faust1961}
\bibinfo{author}{W.~L. Faust}, \bibinfo{author}{M.~N. McDermott},
\newblock \bibinfo{title}{Hyperfine structure of the
  ${(5p)}^{5}(6s)^{3}${P}$_{2}$ state of $_{54}${Xe}$^{129}$ and
  $_{54}${Xe}$^{131}$},
\newblock \bibinfo{journal}{Phys. Rev.} \bibinfo{volume}{123}
  (\bibinfo{year}{1961}) \bibinfo{pages}{198--204}.
  \DOIprefix\doi{10.1103/PhysRev.123.198}.
\bibitem[{Lewty et~al.(2013{\natexlab{a}})Lewty, Chuah, Cazan, Barrett, and
  Sahoo}]{lewty2013}
\bibinfo{author}{N.~C. Lewty}, \bibinfo{author}{B.~L. Chuah},
  \bibinfo{author}{R.~Cazan}, \bibinfo{author}{M.~D. Barrett},
  \bibinfo{author}{B.~K. Sahoo},
\newblock \bibinfo{title}{Experimental determination of the nuclear magnetic
  octupole moment of ${}^{137}$ba${}^{+}$ ion},
\newblock \bibinfo{journal}{Phys. Rev. A} \bibinfo{volume}{88}
  (\bibinfo{year}{2013}{\natexlab{a}}) \bibinfo{pages}{012518}.
  \DOIprefix\doi{10.1103/PhysRevA.88.012518}.
\bibitem[{Lewty et~al.(2013{\natexlab{b}})Lewty, Chuah, Cazan, Sahoo, and
  Barrett}]{lewty2012corrected}
\bibinfo{author}{N.~C. Lewty}, \bibinfo{author}{B.~L. Chuah},
  \bibinfo{author}{R.~Cazan}, \bibinfo{author}{B.~K. Sahoo},
  \bibinfo{author}{M.~D. Barrett},
\newblock \bibinfo{title}{Spectroscopy on a single trapped $^{137}${Ba}+ ion
  for nuclear magnetic octupole moment determination: erratum},
\newblock \bibinfo{journal}{Opt. Express} \bibinfo{volume}{21}
  (\bibinfo{year}{2013}{\natexlab{b}}) \bibinfo{pages}{7131--7132}.
  \DOIprefix\doi{10.1364/OE.21.007131}.
\bibitem[{Hoffman(2014)}]{Hoffman2014}
\bibinfo{author}{M.~R. Hoffman}, \bibinfo{title}{Observation of the Nuclear
  Magnetic Octupole Moment of $^{137}${Ba}+}, Ph.D. thesis, Washington
  University, \bibinfo{year}{2014}. \URLprefix
  \url{http://hdl.handle.net/1773/27558}.
\bibitem[{Unsworth(1969)}]{Unsworth1969}
\bibinfo{author}{P.~J. Unsworth},
\newblock \bibinfo{title}{Nuclear dipole, quadrupole and octupole moments of
  $^{155}\mathrm{Gd}$ by atomic beam magnetic resonance},
\newblock \bibinfo{journal}{J. Phys. B} \bibinfo{volume}{2}
  (\bibinfo{year}{1969}) \bibinfo{pages}{122--133}.
  \DOIprefix\doi{10.1088/0022-3700/2/1/318}.
\bibitem[{Singh et~al.(2013)Singh, Angom, and Natarajan}]{singh2013}
\bibinfo{author}{A.~K. Singh}, \bibinfo{author}{D.~Angom},
  \bibinfo{author}{V.~Natarajan},
\newblock \bibinfo{title}{Observation of the nuclear magnetic octupole moment
  of $^{173}${Yb} from precise measurements of the hyperfine structure in the
  $^{3}${P}$_2$ state},
\newblock \bibinfo{journal}{Phys. Rev. A} \bibinfo{volume}{87}
  (\bibinfo{year}{2013}) \bibinfo{pages}{012512}.
  \DOIprefix\doi{10.1103/PhysRevA.87.012512}.
\bibitem[{de~Groote et~al.(2021)de~Groote, Kujanp\"a\"a, Koszor\'us, Li, and
  Moore}]{deGroote2021}
\bibinfo{author}{R.~P. de~Groote}, \bibinfo{author}{S.~Kujanp\"a\"a},
  \bibinfo{author}{A.~Koszor\'us}, \bibinfo{author}{J.~G. Li},
  \bibinfo{author}{I.~D. Moore},
\newblock \bibinfo{title}{Magnetic octupole moment of $^{173}\mathrm{Yb}$ using
  collinear laser spectroscopy},
\newblock \bibinfo{journal}{Phys. Rev. A} \bibinfo{volume}{103}
  (\bibinfo{year}{2021}) \bibinfo{pages}{032826}.
  \DOIprefix\doi{10.1103/PhysRevA.103.032826}.
\bibitem[{McDermott and Lichten(1960)}]{mcdermott1960}
\bibinfo{author}{M.~N. McDermott}, \bibinfo{author}{W.~L. Lichten},
\newblock \bibinfo{title}{Hyperfine structure of the $6^{3}{P}_{2}$ state of
  $_{80}\mathrm{Hg}^{199}$ and $_{80}\mathrm{Hg}^{201}$. {P}roperties of
  metastable states of mercury},
\newblock \bibinfo{journal}{Phys. Rev.} \bibinfo{volume}{119}
  (\bibinfo{year}{1960}) \bibinfo{pages}{134--143}.
  \DOIprefix\doi{10.1103/PhysRev.119.134}.
\bibitem[{Olsmats et~al.(1961)Olsmats, Axensten, and Liljegren}]{olsmats1961}
\bibinfo{author}{C.~M. Olsmats}, \bibinfo{author}{S.~Axensten},
  \bibinfo{author}{G.~Liljegren},
\newblock \bibinfo{title}{Hyperfine structure investigation of {Po}-205 and
  {Po}-207},
\newblock \bibinfo{journal}{Arkiv for Fysik} \bibinfo{volume}{19}
  (\bibinfo{year}{1961}) \bibinfo{pages}{469--481}.
\bibitem[{Daly and Holloway(1954)}]{daly1954}
\bibinfo{author}{R.~T. Daly}, \bibinfo{author}{J.~H. Holloway},
\newblock \bibinfo{title}{Nuclear magnetic octupole moments of the stable
  {G}allium isotopes},
\newblock \bibinfo{journal}{Phys. Rev.} \bibinfo{volume}{96}
  (\bibinfo{year}{1954}) \bibinfo{pages}{539--540}.
  \DOIprefix\doi{10.1103/PhysRev.96.539}.
\bibitem[{Jaccarino et~al.(1954)Jaccarino, King, Satten, and
  Stroke}]{jaccarino1954}
\bibinfo{author}{V.~Jaccarino}, \bibinfo{author}{J.~G. King},
  \bibinfo{author}{R.~A. Satten}, \bibinfo{author}{H.~H. Stroke},
\newblock \bibinfo{title}{Hyperfine structure of {I}$^{127}$. nuclear magnetic
  octupole moment},
\newblock \bibinfo{journal}{Phys. Rev.} \bibinfo{volume}{94}
  (\bibinfo{year}{1954}) \bibinfo{pages}{1798--1799}.
  \DOIprefix\doi{10.1103/PhysRev.94.1798}.
\bibitem[{Hull and Brink(1970)}]{hull1970}
\bibinfo{author}{R.~J. Hull}, \bibinfo{author}{G.~O. Brink},
\newblock \bibinfo{title}{Hyperfine structure of {Bi}$^{209}$},
\newblock \bibinfo{journal}{Phys. Rev. A} \bibinfo{volume}{1}
  (\bibinfo{year}{1970}) \bibinfo{pages}{685--693}.
  \DOIprefix\doi{10.1103/PhysRevA.1.685}.
\bibitem[{de~Groote et~al.(2022)de~Groote, Moreno, Dobaczewski, Koszor{\'u}s,
  Moore, Reponen, Sahoo, and Yuan}]{deGroote2022}
\bibinfo{author}{R.~P. de~Groote}, \bibinfo{author}{J.~Moreno},
  \bibinfo{author}{J.~Dobaczewski}, \bibinfo{author}{{\'A}.~Koszor{\'u}s},
  \bibinfo{author}{I.~Moore}, \bibinfo{author}{M.~Reponen},
  \bibinfo{author}{B.~Sahoo}, \bibinfo{author}{C.~Yuan},
\newblock \bibinfo{title}{Precision measurement of the magnetic octupole moment
  in 45sc as a test for state-of-the-art atomic-and nuclear-structure theory},
\newblock \bibinfo{journal}{Physics Letters B} \bibinfo{volume}{827}
  (\bibinfo{year}{2022}) \bibinfo{pages}{136930}.
\bibitem[{Ros{\'{e}}n(1972)}]{rosen1972}
\bibinfo{author}{A.~Ros{\'{e}}n},
\newblock \bibinfo{title}{Hyperfine structure analysis for the ground
  configuration of bismuth},
\newblock \bibinfo{journal}{Phys. Scr.} \bibinfo{volume}{6}
  (\bibinfo{year}{1972}) \bibinfo{pages}{37--46}.
  \DOIprefix\doi{10.1088/0031-8949/6/1/004}.
\bibitem[{Suekane and Yamaguchi(1957)}]{Suekane1957}
\bibinfo{author}{S.~Suekane}, \bibinfo{author}{Y.~Yamaguchi},
\newblock \bibinfo{title}{{The Magnetic Octupole Moments of Nuclei}},
\newblock \bibinfo{journal}{Prog. Theor. Phys.} \bibinfo{volume}{17}
  (\bibinfo{year}{1957}) \bibinfo{pages}{443--448}.
  \DOIprefix\doi{10.1143/PTP.17.443}.
\bibitem[{Schwartz(1955)}]{schwartz1955}
\bibinfo{author}{C.~Schwartz},
\newblock \bibinfo{title}{Theory of hyperfine structure},
\newblock \bibinfo{journal}{Phys. Rev.} \bibinfo{volume}{97}
  (\bibinfo{year}{1955}) \bibinfo{pages}{380--395}.
  \DOIprefix\doi{10.1103/PhysRev.97.380}.
\bibitem[{Zhu et~al.(1995)Zhu, Lu, Hamilton, Ramayya, Peker, Wang, Ma, Babu,
  Ginter, Kormicki, Shi, Deng, Nazarewicz, Rasmussen, Stoyer, Chu, Gregorich,
  Mohar, Asztalos, Prussin, Cole, Aryaeinejad, Dardenne, Drigert, Moody,
  Loughed, Wild, Johnson, Lee, McGowan, Ter-Akopian, and Oganessian}]{Zhu1995}
\bibinfo{author}{S.~J. Zhu}, \bibinfo{author}{Q.~H. Lu}, \bibinfo{author}{J.~H.
  Hamilton}, \bibinfo{author}{A.~V. Ramayya}, \bibinfo{author}{L.~K. Peker},
  \bibinfo{author}{M.~G. Wang}, \bibinfo{author}{W.~C. Ma},
  \bibinfo{author}{B.~R.~S. Babu}, \bibinfo{author}{T.~N. Ginter},
  \bibinfo{author}{J.~Kormicki}, \bibinfo{author}{D.~Shi},
  \bibinfo{author}{J.~K. Deng}, \bibinfo{author}{W.~Nazarewicz},
  \bibinfo{author}{J.~O. Rasmussen}, \bibinfo{author}{M.~A. Stoyer},
  \bibinfo{author}{S.~Y. Chu}, \bibinfo{author}{K.~E. Gregorich},
  \bibinfo{author}{M.~F. Mohar}, \bibinfo{author}{S.~Asztalos},
  \bibinfo{author}{S.~G. Prussin}, \bibinfo{author}{J.~D. Cole},
  \bibinfo{author}{R.~Aryaeinejad}, \bibinfo{author}{Y.~K. Dardenne},
  \bibinfo{author}{M.~Drigert}, \bibinfo{author}{K.~J. Moody},
  \bibinfo{author}{R.~W. Loughed}, \bibinfo{author}{J.~F. Wild},
  \bibinfo{author}{N.~R. Johnson}, \bibinfo{author}{I.~Y. Lee},
  \bibinfo{author}{F.~K. McGowan}, \bibinfo{author}{G.~M. Ter-Akopian},
  \bibinfo{author}{Y.~T. Oganessian},
\newblock \bibinfo{title}{Octupole deformation in $^{142,143}$ba and
  $^{144}$ce: new band structures in neutron-rich ba-isotopes},
\newblock \bibinfo{journal}{Phys. Lett. B} \bibinfo{volume}{357}
  (\bibinfo{year}{1995}) \bibinfo{pages}{273 -- 280}.
  \DOIprefix\doi{10.1016/0370-2693(95)00900-6}.
\bibitem[{Auerbach et~al.(1996)Auerbach, Flambaum, and Spevak}]{Auerbach1996}
\bibinfo{author}{N.~Auerbach}, \bibinfo{author}{V.~V. Flambaum},
  \bibinfo{author}{V.~Spevak},
\newblock \bibinfo{title}{Collective t- and p-odd electromagnetic moments in
  nuclei with octupole deformations},
\newblock \bibinfo{journal}{Phys. Rev. Lett.} \bibinfo{volume}{76}
  (\bibinfo{year}{1996}) \bibinfo{pages}{4316--4319}.
  \DOIprefix\doi{10.1103/PhysRevLett.76.4316}.
\bibitem[{Bonatsos et~al.(2005)Bonatsos, Lenis, Minkov, Petrellis, and
  Yotov}]{Bonatsos2005}
\bibinfo{author}{D.~Bonatsos}, \bibinfo{author}{D.~Lenis},
  \bibinfo{author}{N.~Minkov}, \bibinfo{author}{D.~Petrellis},
  \bibinfo{author}{P.~Yotov},
\newblock \bibinfo{title}{Analytic description of critical-point actinides in a
  transition from octupole deformation to octupole vibrations},
\newblock \bibinfo{journal}{Phys. Rev. C} \bibinfo{volume}{71}
  (\bibinfo{year}{2005}) \bibinfo{pages}{064309}.
  \DOIprefix\doi{10.1103/PhysRevC.71.064309}.
\bibitem[{Minkov et~al.(2006)Minkov, Yotov, Drenska, Scheid, Bonatsos, Lenis,
  and Petrellis}]{Minkov2006}
\bibinfo{author}{N.~Minkov}, \bibinfo{author}{P.~Yotov},
  \bibinfo{author}{S.~Drenska}, \bibinfo{author}{W.~Scheid},
  \bibinfo{author}{D.~Bonatsos}, \bibinfo{author}{D.~Lenis},
  \bibinfo{author}{D.~Petrellis},
\newblock \bibinfo{title}{Nuclear collective motion with a coherent coupling
  interaction between quadrupole and octupole modes},
\newblock \bibinfo{journal}{Phys. Rev. C} \bibinfo{volume}{73}
  (\bibinfo{year}{2006}) \bibinfo{pages}{044315}.
  \DOIprefix\doi{10.1103/PhysRevC.73.044315}.
\bibitem[{Robledo and Bertsch(2011)}]{robledo2011}
\bibinfo{author}{L.~M. Robledo}, \bibinfo{author}{G.~F. Bertsch},
\newblock \bibinfo{title}{Global systematics of octupole excitations in
  even-even nuclei},
\newblock \bibinfo{journal}{Physical Review C} \bibinfo{volume}{84}
  (\bibinfo{year}{2011}) \bibinfo{pages}{054302}.
\bibitem[{Gaffney et~al.(2013)Gaffney, Butler, Scheck, Hayes, Wenander, Albers,
  Bastin, Bauer, Blazhev, B{\"o}nig, Bree, Cederk{\"a}ll, Chupp, Cline,
  Cocolios, Davinson, De~Witte, Diriken, Grahn, Herzan, Huyse, Jenkins, Joss,
  Kesteloot, Konki, Kowalczyk, Kr{\"o}ll, Kwan, Lutter, Moschner, Napiorkowski,
  Pakarinen, Pfeiffer, Radeck, Reiter, Reynders, Rigby, Robledo, Rudigier,
  Sambi, Seidlitz, Siebeck, Stora, Thoele, Van~Duppen, Vermeulen, von Schmid,
  Voulot, Warr, Wimmer, Wrzosek-Lipska, Wu, and Zielinska}]{Gaffney2013}
\bibinfo{author}{L.~P. Gaffney}, \bibinfo{author}{P.~A. Butler},
  \bibinfo{author}{M.~Scheck}, \bibinfo{author}{A.~B. Hayes},
  \bibinfo{author}{F.~Wenander}, \bibinfo{author}{M.~Albers},
  \bibinfo{author}{B.~Bastin}, \bibinfo{author}{C.~Bauer},
  \bibinfo{author}{A.~Blazhev}, \bibinfo{author}{S.~B{\"o}nig},
  \bibinfo{author}{N.~Bree}, \bibinfo{author}{J.~Cederk{\"a}ll},
  \bibinfo{author}{T.~Chupp}, \bibinfo{author}{D.~Cline},
  \bibinfo{author}{T.~E. Cocolios}, \bibinfo{author}{T.~Davinson},
  \bibinfo{author}{H.~De~Witte}, \bibinfo{author}{J.~Diriken},
  \bibinfo{author}{T.~Grahn}, \bibinfo{author}{A.~Herzan},
  \bibinfo{author}{M.~Huyse}, \bibinfo{author}{D.~G. Jenkins},
  \bibinfo{author}{D.~T. Joss}, \bibinfo{author}{N.~Kesteloot},
  \bibinfo{author}{J.~Konki}, \bibinfo{author}{M.~Kowalczyk},
  \bibinfo{author}{T.~Kr{\"o}ll}, \bibinfo{author}{E.~Kwan},
  \bibinfo{author}{R.~Lutter}, \bibinfo{author}{K.~Moschner},
  \bibinfo{author}{P.~Napiorkowski}, \bibinfo{author}{J.~Pakarinen},
  \bibinfo{author}{M.~Pfeiffer}, \bibinfo{author}{D.~Radeck},
  \bibinfo{author}{P.~Reiter}, \bibinfo{author}{K.~Reynders},
  \bibinfo{author}{S.~V. Rigby}, \bibinfo{author}{L.~M. Robledo},
  \bibinfo{author}{M.~Rudigier}, \bibinfo{author}{S.~Sambi},
  \bibinfo{author}{M.~Seidlitz}, \bibinfo{author}{B.~Siebeck},
  \bibinfo{author}{T.~Stora}, \bibinfo{author}{P.~Thoele},
  \bibinfo{author}{P.~Van~Duppen}, \bibinfo{author}{M.~J. Vermeulen},
  \bibinfo{author}{M.~von Schmid}, \bibinfo{author}{D.~Voulot},
  \bibinfo{author}{N.~Warr}, \bibinfo{author}{K.~Wimmer},
  \bibinfo{author}{K.~Wrzosek-Lipska}, \bibinfo{author}{C.~Y. Wu},
  \bibinfo{author}{M.~Zielinska},
\newblock \bibinfo{title}{Studies of pear-shaped nuclei using accelerated
  radioactive beams},
\newblock \bibinfo{journal}{Nature} \bibinfo{volume}{497}
  (\bibinfo{year}{2013}) \bibinfo{pages}{199--204}.
  \DOIprefix\doi{10.1038/nature12073}.
\bibitem[{Bonatsos et~al.(2015)Bonatsos, Martinou, Minkov, Karampagia, and
  Petrellis}]{Bonatsos2015}
\bibinfo{author}{D.~Bonatsos}, \bibinfo{author}{A.~Martinou},
  \bibinfo{author}{N.~Minkov}, \bibinfo{author}{S.~Karampagia},
  \bibinfo{author}{D.~Petrellis},
\newblock \bibinfo{title}{Octupole deformation in light actinides within an
  analytic quadrupole octupole axially symmetric model with a davidson
  potential},
\newblock \bibinfo{journal}{Phys. Rev. C} \bibinfo{volume}{91}
  (\bibinfo{year}{2015}) \bibinfo{pages}{054315}.
  \DOIprefix\doi{10.1103/PhysRevC.91.054315}.
\bibitem[{Brewer et~al.(2018)Brewer, Wang, Yzaguirre, Hamilton, Ramayya, Liu,
  Hwang, Luo, Zachary, Rasmussen, Zhu, Goodin, Ter-Akopian, Daniel, and
  Oganessian}]{Brewer2017}
\bibinfo{author}{N.~T. Brewer}, \bibinfo{author}{E.~H. Wang},
  \bibinfo{author}{W.~A. Yzaguirre}, \bibinfo{author}{J.~H. Hamilton},
  \bibinfo{author}{A.~V. Ramayya}, \bibinfo{author}{S.~H. Liu},
  \bibinfo{author}{J.~K. Hwang}, \bibinfo{author}{Y.~X. Luo},
  \bibinfo{author}{C.~J. Zachary}, \bibinfo{author}{J.~O. Rasmussen},
  \bibinfo{author}{S.~J. Zhu}, \bibinfo{author}{C.~Goodin},
  \bibinfo{author}{G.~M. Ter-Akopian}, \bibinfo{author}{A.~V. Daniel},
  \bibinfo{author}{Y.~T. Oganessian}, \bibinfo{title}{Octupole Deformation in
  $^{144}${Ba} and $^{148}${Ce}}, \bibinfo{year}{2018}, pp.
  \bibinfo{pages}{197--202}. \DOIprefix\doi{10.1142/9789813229426_0038}.
\bibitem[{Butler et~al.(2020)Butler, Gaffney, Spagnoletti, Abrahams, Bowry,
  Cederk\"all, de~Angelis, De~Witte, Garrett, Goldkuhle, Henrich, Illana,
  Johnston, Joss, Keatings, Kelly, Komorowska, Konki, Kr\"oll, Lozano,
  Nara~Singh, O'Donnell, Ojala, Page, Pedersen, Raison, Reiter, Rodriguez,
  Rosiak, Rothe, Scheck, Seidlitz, Shneidman, Siebeck, Sinclair, Smith,
  Stryjczyk, Van~Duppen, Vinals, Virtanen, Warr, Wrzosek-Lipska, and
  Zieli\ifmmode~\acute{n}\else \'{n}\fi{}ska}]{Butler2020}
\bibinfo{author}{P.~A. Butler}, \bibinfo{author}{L.~P. Gaffney},
  \bibinfo{author}{P.~Spagnoletti}, \bibinfo{author}{K.~Abrahams},
  \bibinfo{author}{M.~Bowry}, \bibinfo{author}{J.~Cederk\"all},
  \bibinfo{author}{G.~de~Angelis}, \bibinfo{author}{H.~De~Witte},
  \bibinfo{author}{P.~E. Garrett}, \bibinfo{author}{A.~Goldkuhle},
  \bibinfo{author}{C.~Henrich}, \bibinfo{author}{A.~Illana},
  \bibinfo{author}{K.~Johnston}, \bibinfo{author}{D.~T. Joss},
  \bibinfo{author}{J.~M. Keatings}, \bibinfo{author}{N.~A. Kelly},
  \bibinfo{author}{M.~Komorowska}, \bibinfo{author}{J.~Konki},
  \bibinfo{author}{T.~Kr\"oll}, \bibinfo{author}{M.~Lozano},
  \bibinfo{author}{B.~S. Nara~Singh}, \bibinfo{author}{D.~O'Donnell},
  \bibinfo{author}{J.~Ojala}, \bibinfo{author}{R.~D. Page},
  \bibinfo{author}{L.~G. Pedersen}, \bibinfo{author}{C.~Raison},
  \bibinfo{author}{P.~Reiter}, \bibinfo{author}{J.~A. Rodriguez},
  \bibinfo{author}{D.~Rosiak}, \bibinfo{author}{S.~Rothe},
  \bibinfo{author}{M.~Scheck}, \bibinfo{author}{M.~Seidlitz},
  \bibinfo{author}{T.~M. Shneidman}, \bibinfo{author}{B.~Siebeck},
  \bibinfo{author}{J.~Sinclair}, \bibinfo{author}{J.~F. Smith},
  \bibinfo{author}{M.~Stryjczyk}, \bibinfo{author}{P.~Van~Duppen},
  \bibinfo{author}{S.~Vinals}, \bibinfo{author}{V.~Virtanen},
  \bibinfo{author}{N.~Warr}, \bibinfo{author}{K.~Wrzosek-Lipska},
  \bibinfo{author}{M.~Zieli\ifmmode~\acute{n}\else \'{n}\fi{}ska},
\newblock \bibinfo{title}{Evolution of octupole deformation in radium nuclei
  from coulomb excitation of radioactive $^{222}${Ra} and $^{228}${Ra} beams},
\newblock \bibinfo{journal}{Phys. Rev. Lett.} \bibinfo{volume}{124}
  (\bibinfo{year}{2020}) \bibinfo{pages}{042503}.
  \DOIprefix\doi{10.1103/PhysRevLett.124.042503}.
\bibitem[{Chinn et~al.(1996)Chinn, Umar, Vallieres, and Strayer}]{chinn1996}
\bibinfo{author}{C.~R. Chinn}, \bibinfo{author}{A.~S. Umar},
  \bibinfo{author}{M.~Vallieres}, \bibinfo{author}{M.~R. Strayer},
\newblock \bibinfo{title}{Mean field studies of exotic nuclei},
\newblock \bibinfo{journal}{Physics reports} \bibinfo{volume}{264}
  (\bibinfo{year}{1996}) \bibinfo{pages}{107--121}.
\bibitem[{Sen'kov and Dmitriev(2002)}]{senkov2002}
\bibinfo{author}{R.~A. Sen'kov}, \bibinfo{author}{V.~F. Dmitriev},
\newblock \bibinfo{title}{Nuclear magnetization distribution and hyperfine
  splitting in $\mathrm{Bi}^{82+}$ ion},
\newblock \bibinfo{journal}{Nucl. Phys. A} \bibinfo{volume}{706}
  (\bibinfo{year}{2002}) \bibinfo{pages}{351 -- 364}.
  \DOIprefix\doi{10.1016/S0375-9474(02)00759-5}.
\bibitem[{Mertzimekis(2016)}]{mertzimekis2016}
\bibinfo{author}{T.~J. Mertzimekis},
\newblock \bibinfo{title}{Prediction and evaluation of magnetic moments in
  $t=1/2$, 3/2, and 5/2 mirror nuclei},
\newblock \bibinfo{journal}{Phys. Rev. C} \bibinfo{volume}{94}
  (\bibinfo{year}{2016}) \bibinfo{pages}{064313}.
  \DOIprefix\doi{10.1103/PhysRevC.94.064313}.
\bibitem[{Bohr and Mottelson(1953)}]{Bohr-Mottelson}
\bibinfo{author}{A.~N. Bohr}, \bibinfo{author}{B.~R. Mottelson},
\newblock \bibinfo{title}{Collective and individual-particle aspects of nuclear
  structure},
\newblock \bibinfo{journal}{Dan. Mat. Fys. Medd.} \bibinfo{volume}{27}
  (\bibinfo{year}{1953}) \bibinfo{pages}{1--174}.
\bibitem[{Rosenfeld(1948)}]{rosenfeld1948}
\bibinfo{author}{L.~Rosenfeld}, \bibinfo{title}{Nuclear Forces},
  \bibinfo{publisher}{North-Holland}, \bibinfo{year}{1948}.
\bibitem[{Fuller(1976)}]{fuller1976}
\bibinfo{author}{G.~H. Fuller},
\newblock \bibinfo{title}{Nuclear spins and moments},
\newblock \bibinfo{journal}{J. Phys. Chem. Ref. Data} \bibinfo{volume}{5}
  (\bibinfo{year}{1976}) \bibinfo{pages}{835--1092}.
  \DOIprefix\doi{10.1063/1.555544}.
\bibitem[{Mertzimekis et~al.(2016)Mertzimekis, Stamou, and
  Psaltis}]{mertzimekis2016NUMOR}
\bibinfo{author}{T.~J. Mertzimekis}, \bibinfo{author}{K.~Stamou},
  \bibinfo{author}{A.~Psaltis},
\newblock \bibinfo{title}{An online database of nuclear electromagnetic
  moments},
\newblock \bibinfo{journal}{Nucl. Instrum. Meth. Phys. Res. A}
  \bibinfo{volume}{807} (\bibinfo{year}{2016}) \bibinfo{pages}{56--60}.
  \DOIprefix\doi{10.1016/j.nima.2015.10.096}.
\bibitem[{Mertzimekis et~al.(2024)Mertzimekis, Pelonis, and
  Papanastasopoulos}]{2024_Mertzimekis}
\bibinfo{author}{T.~J. Mertzimekis}, \bibinfo{author}{S.~Pelonis},
  \bibinfo{author}{K.~Papanastasopoulos},
\newblock \bibinfo{title}{An upgrade of the nuclear moments database {NUMOR}}
  (\bibinfo{year}{2024}). \bibinfo{note}{~(in prep.)}.
\bibitem[{Seth et~al.(1958)Seth, Hughes, Zimmerman, and Garth}]{seth1958}
\bibinfo{author}{K.~K. Seth}, \bibinfo{author}{D.~J. Hughes},
  \bibinfo{author}{R.~L. Zimmerman}, \bibinfo{author}{R.~C. Garth},
\newblock \bibinfo{title}{Nuclear radii by scattering of low-energy neutrons},
\newblock \bibinfo{journal}{Phys. Rev.} \bibinfo{volume}{110}
  (\bibinfo{year}{1958}) \bibinfo{pages}{692--700}.
  \DOIprefix\doi{10.1103/PhysRev.110.692}.
\bibitem[{Casten et~al.(1993)Casten, Zamfir, and Brenner}]{casten1993}
\bibinfo{author}{R.~F. Casten}, \bibinfo{author}{N.~V. Zamfir},
  \bibinfo{author}{D.~S. Brenner},
\newblock \bibinfo{title}{A new simple global phenomenology of nuclear shape
  transitions},
\newblock in: \bibinfo{booktitle}{Proceedings of the International Conference
  on radioactive nuclear beams}, \bibinfo{year}{24-27 May 1993;}. \URLprefix
  \url{https://www.osti.gov/biblio/10172873}.
\bibitem[{Brown et~al.(1980)Brown, Chung, and Wildenthal}]{brown1980}
\bibinfo{author}{B.~A. Brown}, \bibinfo{author}{W.~Chung},
  \bibinfo{author}{B.~H. Wildenthal},
\newblock \bibinfo{title}{Electromagnetic multipole moments of ground states of
  stable odd-mass nuclei in the sd shell},
\newblock \bibinfo{journal}{Physical Review C} \bibinfo{volume}{22}
  (\bibinfo{year}{1980}) \bibinfo{pages}{774}.
\bibitem[{Noya et~al.(1958)Noya, Arima, and Horie}]{noya1958}
\bibinfo{author}{H.~Noya}, \bibinfo{author}{A.~Arima},
  \bibinfo{author}{H.~Horie},
\newblock \bibinfo{title}{Nuclear moments and configuration mixing},
\newblock \bibinfo{journal}{Progress of Theoretical Physics Supplement}
  \bibinfo{volume}{8} (\bibinfo{year}{1958}) \bibinfo{pages}{33--112}.
\bibitem[{Van~Hees and Glaudemans(1981)}]{van1981}
\bibinfo{author}{A.~G.~M. Van~Hees}, \bibinfo{author}{P.~W.~M. Glaudemans},
\newblock \bibinfo{title}{Effects of configuration mixing on predominantlyf 7
  2/n states ina= 52--55 nuclei},
\newblock \bibinfo{journal}{Zeitschrift f{\"u}r Physik A Atoms and Nuclei}
  \bibinfo{volume}{303} (\bibinfo{year}{1981}) \bibinfo{pages}{267--276}.
\bibitem[{Wolf and Casten(1987)}]{wolf1987}
\bibinfo{author}{A.~Wolf}, \bibinfo{author}{R.~F. Casten},
\newblock \bibinfo{title}{Effective valence proton and neutron numbers in
  transitional $a\cong 150$ nuclei from b(e2) and g-factor data},
\newblock \bibinfo{journal}{Physical Review C} \bibinfo{volume}{36}
  (\bibinfo{year}{1987}) \bibinfo{pages}{851}.
\bibitem[{Vergados(1971)}]{vergados1971}
\bibinfo{author}{J.~D. Vergados},
\newblock \bibinfo{title}{Configuration mixing effects on m1 transitions and
  magnetic moments in pb region},
\newblock \bibinfo{journal}{Physics Letters B} \bibinfo{volume}{36}
  (\bibinfo{year}{1971}) \bibinfo{pages}{12--17}.
\bibitem[{Mertzimekis et~al.(2003)Mertzimekis, Stuchbery, Benczer-Koller, and
  Taylor}]{mertzimekis2003}
\bibinfo{author}{T.~J. Mertzimekis}, \bibinfo{author}{A.~E. Stuchbery},
  \bibinfo{author}{N.~Benczer-Koller}, \bibinfo{author}{M.~J. Taylor},
\newblock \bibinfo{title}{Systematics of first 2+ state g factors around mass
  80},
\newblock \bibinfo{journal}{Physical Review C} \bibinfo{volume}{68}
  (\bibinfo{year}{2003}) \bibinfo{pages}{054304}.
\bibitem[{Stuchbery et~al.(1991)Stuchbery, Lampard, and
  Bolotin}]{stuchbery1991}
\bibinfo{author}{A.~E. Stuchbery}, \bibinfo{author}{G.~J. Lampard},
  \bibinfo{author}{H.~H. Bolotin},
\newblock \bibinfo{title}{Transient field measurements of g-factors in
  194,196,198 pt; g (21+) systematics in transitional w, os, pt nuclei},
\newblock \bibinfo{journal}{Nuclear Physics A} \bibinfo{volume}{528}
  (\bibinfo{year}{1991}) \bibinfo{pages}{447--464}.
\bibitem[{Buck and Perez(1983)}]{buck1983}
\bibinfo{author}{B.~Buck}, \bibinfo{author}{S.~M. Perez},
\newblock \bibinfo{title}{New look at magnetic moments and beta decays of
  mirror nuclei},
\newblock \bibinfo{journal}{Phys. Rev. Lett.} \bibinfo{volume}{50}
  (\bibinfo{year}{1983}) \bibinfo{pages}{1975--1978}.
  \DOIprefix\doi{10.1103/PhysRevLett.50.1975}.
\bibitem[{Buck et~al.(2001)Buck, Merchant, and Perez}]{buck2001}
\bibinfo{author}{B.~Buck}, \bibinfo{author}{A.~C. Merchant},
  \bibinfo{author}{S.~M. Perez},
\newblock \bibinfo{title}{Magnetic moments of mirror nuclei},
\newblock \bibinfo{journal}{Phys. Rev. C} \bibinfo{volume}{63}
  (\bibinfo{year}{2001}) \bibinfo{pages}{037301}.
  \DOIprefix\doi{10.1103/PhysRevC.63.037301}.
\bibitem[{Hanna and Hugg(1985)}]{hanna1985}
\bibinfo{author}{S.~S. Hanna}, \bibinfo{author}{J.~W. Hugg},
\newblock \bibinfo{title}{Analysis of the magnetic moments of mirror nuclei},
\newblock \bibinfo{journal}{Hyperfine Interact.} \bibinfo{volume}{21}
  (\bibinfo{year}{1985}) \bibinfo{pages}{59--77}.
  \DOIprefix\doi{10.1007/BF02061977}.
\bibitem[{Mertzimekis et~al.(2006)Mertzimekis, Mantica, Davies, Liddick, and
  Tomlin}]{mertzimekis2006}
\bibinfo{author}{T.~J. Mertzimekis}, \bibinfo{author}{P.~F. Mantica},
  \bibinfo{author}{A.~D. Davies}, \bibinfo{author}{S.~N. Liddick},
  \bibinfo{author}{B.~E. Tomlin},
\newblock \bibinfo{title}{Ground state magnetic dipole moment of
  $^{35}\mathrm{K}$},
\newblock \bibinfo{journal}{Phys. Rev. C} \bibinfo{volume}{73}
  (\bibinfo{year}{2006}) \bibinfo{pages}{024318}.
  \DOIprefix\doi{10.1103/PhysRevC.73.024318}.
\bibitem[{Berryman et~al.(2009)Berryman, Minamisono, Rogers, Brown, Crawford,
  Grinyer, Mantica, Stoker, and Towner}]{berryman2009}
\bibinfo{author}{J.~S. Berryman}, \bibinfo{author}{K.~Minamisono},
  \bibinfo{author}{W.~F. Rogers}, \bibinfo{author}{B.~A. Brown},
  \bibinfo{author}{H.~L. Crawford}, \bibinfo{author}{G.~F. Grinyer},
  \bibinfo{author}{P.~F. Mantica}, \bibinfo{author}{J.~B. Stoker},
  \bibinfo{author}{I.~S. Towner},
\newblock \bibinfo{title}{Doubly-magic nature of $^{56}\mathrm{Ni}$:
  Measurement of the ground state nuclear magnetic dipole moment of
  $^{55}\mathrm{Ni}$},
\newblock \bibinfo{journal}{Phys. Rev. C} \bibinfo{volume}{79}
  (\bibinfo{year}{2009}) \bibinfo{pages}{064305}.
  \DOIprefix\doi{10.1103/PhysRevC.79.064305}.
\bibitem[{Perez et~al.(2008)Perez, Richter, Brown, and Horoi}]{perez2008}
\bibinfo{author}{S.~M. Perez}, \bibinfo{author}{W.~A. Richter},
  \bibinfo{author}{B.~A. Brown}, \bibinfo{author}{M.~Horoi},
\newblock \bibinfo{title}{Correlations between magnetic moments and
  $\ensuremath{\beta}$ decays of mirror nuclei},
\newblock \bibinfo{journal}{Phys. Rev. C} \bibinfo{volume}{77}
  (\bibinfo{year}{2008}) \bibinfo{pages}{064311}.
  \DOIprefix\doi{10.1103/PhysRevC.77.064311}.
\bibitem[{Minamisono et~al.(2006)Minamisono, Mantica, Mertzimekis, Davies,
  Hass, Pereira, Pinter, Rogers, Stoker, Tomlin, and
  Weerasiri}]{minamisono2006}
\bibinfo{author}{K.~Minamisono}, \bibinfo{author}{P.~F. Mantica},
  \bibinfo{author}{T.~J. Mertzimekis}, \bibinfo{author}{A.~D. Davies},
  \bibinfo{author}{M.~Hass}, \bibinfo{author}{J.~Pereira},
  \bibinfo{author}{J.~S. Pinter}, \bibinfo{author}{W.~F. Rogers},
  \bibinfo{author}{J.~B. Stoker}, \bibinfo{author}{B.~E. Tomlin},
  \bibinfo{author}{R.~R. Weerasiri},
\newblock \bibinfo{title}{Nuclear magnetic moment of the $^{57}\mathrm{Cu}$
  ground state},
\newblock \bibinfo{journal}{Phys. Rev. Lett.} \bibinfo{volume}{96}
  (\bibinfo{year}{2006}) \bibinfo{pages}{102501}.
  \DOIprefix\doi{10.1103/PhysRevLett.96.102501}.
\bibitem[{Sugimoto(1969)}]{sugimoto1969}
\bibinfo{author}{K.~Sugimoto},
\newblock \bibinfo{title}{Magnetic moments and $\ensuremath{\beta}$-decay
  $\mathrm{ft}$ values of mirror nuclei},
\newblock \bibinfo{journal}{Phys. Rev.} \bibinfo{volume}{182}
  (\bibinfo{year}{1969}) \bibinfo{pages}{1051--1054}.
  \DOIprefix\doi{10.1103/PhysRev.182.1051}.
\bibitem[{Kondev et~al.(2021)Kondev, Wang, Huang, Naimi, and
  Audi}]{2021_Kondev}
\bibinfo{author}{F.~G. Kondev}, \bibinfo{author}{M.~Wang},
  \bibinfo{author}{W.~J. Huang}, \bibinfo{author}{S.~Naimi},
  \bibinfo{author}{G.~Audi},
\newblock \bibinfo{title}{The nubase2020 evaluation of nuclear physics
  properties},
\newblock \bibinfo{journal}{Chinese Physics C} \bibinfo{volume}{45}
  (\bibinfo{year}{2021}) \bibinfo{pages}{030001}.
  \DOIprefix\doi{10.1088/1674-1137/abddae}.

\end{thebibliography}

\clearpage

\section*{Figures}

\clearpage
\TableExplanation

\section*{Table~1. Comparison between experimental and theoretical values of magnetic octupole moments for odd-neutron nuclei.}

Comparison between experimental ($exp$) and theoretical (single-particle ($s.p.$) and collective strong ($scc$) and weak ($wcc$) coupling) values of magnetic octupole moments for odd-neutron nuclei. The equations used to calculate the various $\Omega$ values are listed in the first row. $\Omega/\langle r^2\rangle$ is expressed in unit of $\mu_N$. The second value quoted for each isotope has been calculated with effective values $g_s^{eff}=0.7g_s$ and $g_l^{eff}=g_l+\delta g_l$, where $\delta g_l=+0.1$ for an unpaired proton and $\delta g_l=0.0$ for an unpaired neutron.

\begin{center}
\begin{tabular}{lll}
Isotope & \multicolumn{2}{l}{Experimental and theoretical values of magnetic octupole moments of odd-neutron isotopes} \\

$I^\pi$ & \multicolumn{2}{l}{Spin and parity of the state (from Ref.~\cite{2021_Kondev})}\\

Shell assign. & \multicolumn{2}{l}{Shows the shell assignment for each isotope }\\

Experimental & \multicolumn{2}{l}{Available experimental data in literature (see next column on the right for the corresponding }\\
        & \multicolumn{2}{l}{published works). Notation is as follows:}\\
        & $(a)$ & This value has been used in Fig.~\ref{fig:Om_Sum_up_neutron}\\
        & $(b)$ & The different sign is a result of a different convention of how the octupole moment is\\
        & & related to the hyperfine $C$ constant\\
        
Ref. & \multicolumn{2}{l}{References to literature with experimental values of magnetic octupole moments}\\

s.p. model & \multicolumn{2}{l}{Single-particle model calculations, $\Omega_{sp}/\langle r^2\rangle$ (via Eq.~\ref{eq:Omega shell model})}\\

Collective model & \multicolumn{2}{l}{Calculated octupole moments with the collective model.}\\
        & \multicolumn{2}{l}{Two calculations involved strong coupling:}\\
        && $\Omega_o/\langle r^2\rangle$ (via Eq.~\ref{eq:Omega0 for scc}),\\
        && $\Omega_{scc}/\langle r^2\rangle$ (via Eqs.~\ref{eq:Omega value strong cc} and~\ref{eq:Omega value strong cc for I=3/2}),\\
        & \multicolumn{2}{l}{while three involved weak coupling:}\\
        && $\Omega_{p}/\langle r^2\rangle$ (via Eq.~\ref{eq:Omega_p value weak cc simplified version})\\
        && $\Omega_{c}/\langle r^2\rangle$ (via Eq.~\ref{eq:Omega_c value weak cc simplified version}), and\\
        && $\Omega_{wcc}/\langle r^2\rangle$ (via Eq.~\ref{eq:Omega break2}), respectively.

\end{tabular}
\end{center}
\clearpage
\datatables 
\LTright=0pt
\LTleft=0pt

\section*{Table~2. Comparison between experimental and theoretical values of magnetic octupole moments for odd-proton nuclei.}

Comparison between experimental ($exp$) and theoretical (single-particle ($s.p.$) and collective strong ($scc$) and weak ($wcc$) coupling) values of magnetic octupole moments for odd-proton nuclei. The equations used to calculate the various $\Omega$ values are listed in the first row. $\Omega/\langle r^2\rangle$ is expressed in unit of $\mu_N$. The second value quoted for each isotope has been calculated with effective values $g_s^{eff}=0.7g_s$ and $g_l^{eff}=g_l+\delta g_l$, where $\delta g_l=+0.1$ for an unpaired proton and $\delta g_l=0.0$ for an unpaired neutron.

\begin{center}
\begin{tabular}{lll}
Isotope & \multicolumn{2}{l}{Experimental and theoretical values of magnetic octupole moments of odd-proton isotopes} \\

$I^\pi$ & \multicolumn{2}{l}{Spin and parity of the state (from Ref.~\cite{2021_Kondev})}\\

Shell assign. & \multicolumn{2}{l}{Shows the shell assignment for each isotope }\\

Experimental & \multicolumn{2}{l}{Available experimental data in literature (see next column on the right for the corresponding }\\
        & \multicolumn{2}{l}{published works). Notation is as follows:}\\
        & $(a)$ & This value has been used in Fig.~\ref{fig:Om_Sum_up_proton}\\
        & $(b)$ & No cited experimental uncertainty in literature\\        
Ref. & \multicolumn{2}{l}{References to literature with experimental values of magnetic octupole moments}\\

s.p. model & \multicolumn{2}{l}{Single-particle model calculations, $\Omega_{sp}/\langle r^2\rangle$ (via Eq.~\ref{eq:Omega shell model})}\\

Collective model & \multicolumn{2}{l}{Calculated octupole moments with the collective model.}\\
        & \multicolumn{2}{l}{Two calculations involved strong coupling:}\\
        && $\Omega_o/\langle r^2\rangle$ (via Eq.~\ref{eq:Omega0 for scc}),\\
        && $\Omega_{scc}/\langle r^2\rangle$ (via Eqs.~\ref{eq:Omega value strong cc} and~\ref{eq:Omega value strong cc for I=3/2}),\\
        & \multicolumn{2}{l}{while three involved weak coupling:}\\
        && $\Omega_{p}/\langle r^2\rangle$ (via Eq.~\ref{eq:Omega_p value weak cc simplified version})\\
        && $\Omega_{c}/\langle r^2\rangle$ (via Eq.~\ref{eq:Omega_c value weak cc simplified version}), and\\
        && $\Omega_{wcc}/\langle r^2\rangle$ (via Eq.~\ref{eq:Omega break2}), respectively.

\end{tabular}
\end{center}
\clearpage

\section*{Table~3. Predictions for magnetic octupole moments for isotopes with valence protons in subshell $2p_{3/2}$}

Predictions for isotopes with a (bare) valence proton in subshell $2p_{3/2}$.
The respective spin/parity ($I^\pi$) is also shown (from Ref.~\cite{2021_Kondev}).
    All values are in $\mu_N$ units. The various $\Omega$ terms
    are calculated using
    Eqs.~\ref{eq:Omega shell model},
    \ref{eq:Omega value strong cc},
    \ref{eq:Omega0 for scc},
    \ref{eq:Omega value strong cc for I=3/2},
    \ref{eq:Omega break2},
    \ref{eq:Omega_c value weak cc simplified version} and
    \ref{eq:Omega_p value weak cc simplified version},
    similarly to Tables~\ref{tab:valence_neutron}
    and \ref{tab:valence_proton}.

\section*{Table~4. Predictions for magnetic octupole moments for isotopes with valence protons in subshell $1g_{9/2}$}

Predictions for isotopes with a (bare) valence proton in subshell $1g_{9/2}$.
The respective spin/parity ($I^\pi$) is also shown (from Ref.~\cite{2021_Kondev}).
    All values are in $\mu_N$ units. The various $\Omega$ terms
    are calculated using
    Eqs.~\ref{eq:Omega shell model},
    \ref{eq:Omega value strong cc},
    \ref{eq:Omega0 for scc},
    \ref{eq:Omega value strong cc for I=3/2},
    \ref{eq:Omega break2},
    \ref{eq:Omega_c value weak cc simplified version} and
    \ref{eq:Omega_p value weak cc simplified version},
    similarly to Tables~\ref{tab:valence_neutron}
    and \ref{tab:valence_proton}.

\section*{Table~5. Predictions for magnetic octupole moments for isotopes with valence neutrons in subshell $1g_{9/2}$}

    Collective model predictions (strong and weak coupling) for isotopes with a (bare) valence neutron in subshell $1g_{9/2}$.
    The respective spin/parity ($I^\pi$) is also shown (from Ref.~\cite{2021_Kondev}).
All values are in $\mu_N$ units.

\clearpage

\scriptsize
\begin{table}[ht]
\caption{Comparison between experimental ($exp$) and theoretical (single-particle ($s.p.$) and collective strong ($scc$) and weak ($wcc$) coupling) values of magnetic octupole moments for odd-neutron nuclei.}
\centering
\begin{tabular}{|l|c|c|c|c|c|c|c|c|c|c|}
\hline
  \multicolumn{1}{|c|}{\multirow{3}{*}{Isotope}}
& \multicolumn{1}{c|} {\multirow{3}{*}{$I^\pi$}}
& \multicolumn{1}{c|} {\multirow{3}{*}{\makecell{Shell\\assign.}}}
& \multicolumn{1}{c|} {\multirow{2}{*}{Experimental}}
& \multicolumn{1}{c|} {\multirow{3}{*}{Ref.}}
& \multicolumn{1}{c|} {\multirow{2}{*}{\makecell{s.p.\\model}}}
& \multicolumn{5}{c|} {Collective model} \\ \cline{7-11}

&
&
&
&
&
& \multicolumn{2}{c|}{{\em Strong coupling}} & \multicolumn{3}{c|}{{\em Weak coupling}} \\ \cline{4-4}\cline{6-11}

&
&
& \multicolumn{1}{c|}{~$\Omega_{exp}/\langle r^2\rangle$} & & \multicolumn{1}{c|}{~$\Omega_{sp}/\langle r^2\rangle$}
& \multicolumn{1}{c|}{~$\Omega_{o}/\langle r^2\rangle$} & \multicolumn{1}{c|}{~$\Omega_{scc}/\langle r^2\rangle$}
& \multicolumn{1}{c|}{~$\Omega_{p}/\langle r^2\rangle$} & \multicolumn{1}{c|}{$\Omega_{c}/\langle r^2\rangle$}
& \multicolumn{1}{c|}{~$\Omega_{wcc}/\langle r^2\rangle$}\\
\hline

    
  \multicolumn{5}{|r|}{Equations used:~~}
& \multicolumn{1}{c|}{(\ref{eq:Omega shell model})}
& \multicolumn{1}{c|}{(\ref{eq:Omega0 for scc})}
& \multicolumn{1}{c|}{(\ref{eq:Omega value strong cc}) (\ref{eq:Omega value strong cc for I=3/2})}
& \multicolumn{1}{c|}{(\ref{eq:Omega_p value weak cc simplified version})}
& \multicolumn{1}{c|}{(\ref{eq:Omega_c value weak cc simplified version})}
& \multicolumn{1}{c|}{(\ref{eq:Omega break2})}\\
\hline
%
%
%
  \ce{^{83}Kr}
& $\frac{9}{2}^+$
& $g_{9/2}$
& $-0.87(29)$
& \cite{faust1963}
& \makecell{$-2.087$\\/$-1.461$}
& $-0.361$
& \makecell{$-0.719$\\/$-0.535$}
& \makecell{$-1.483$\\/$-1.038$}
& $-0.088$ & \makecell{$-1.571$\\/$-1.126$} \\
\hline
  \ce{^{131}Xe}
& $\frac{3}{2}^+$
& $d_{3/2}$
& +0.17(4)
& \cite{faust1961}
& \makecell{$+0.164$\\/$+0.115$}
& $-$0.343
& \makecell{$+0.079$\\/$+0.052$}
& \makecell{$+0.101$\\/$+0.071$}
& $-$0.056
& \makecell{$+0.045$\\/$+0.014$}\\
\hline
  \multirow{5}{*}{\ce{^{137}Ba}$^+$}
& \multirow{5}{*}{$\frac{3}{2}^+$}
& \multirow{5}{*}{$d_{3/2}$}
& +0.1742(19)$^{(a)}$
& \cite{lewty2013}
& \multirow{5}{*}{\makecell{+0.164\\/+0.115}}
& \multirow{5}{*}{$-0.341$}
& \multirow{5}{*}{\makecell{+0.079\\/+0.053}}
& \multirow{5}{*}{\makecell{+0.099\\/+0.070}}
& \multirow{5}{*}{$-$0.058}
& \multirow{5}{*}{\makecell{~~+0.042~~\\/+0.012}}\\ \cline{4-5}
&
&
& +0.1740(18)
&\cite{lewty2013}
&
&
&
&
&
& \\ \cline{4-5}
&
&
& +0.171(13)
& \cite{lewty2013}
&
&
&
&
&
& \\ \cline{4-5}
&
&
& $-$0.1681(18)$^{(b)}$
& \cite{lewty2012corrected}
&
&
&
&
&
& \\ \cline{4-5}
&
&
& +0.217(4)
& \cite{Hoffman2014}
&
&
&
&
&
&\\ \hline  

  \ce{^{155}Gd}
& $\frac{3}{2}^-$
& $p_{3/2}$
& $-$5.1(1.9)
& \cite{Unsworth1969}
& \makecell{$-$1.148\\/$-$0.803}
& $-$0.344 & \makecell{$-$0.633\\/$-$0.446}
& \makecell{$-$0.592\\/$-$0.414}
& $-$0.071 & \makecell{$-$0.663\\/$-$0.486} \\
\hline

  \multirow{2}{*}{\ce{^{173}Yb}}
& \multirow{2}{*}{$\frac{5}{2}^-$}
& \multirow{2}{*}{$f_{5/2}$}
& $-$101(6)
& \cite{singh2013}
& \multirow{2}{*}{\makecell{+0.547\\/+0.383}}
& \multirow{2}{*}{$-$0.337}
& \multirow{2}{*}{\makecell{+0.025\\/+0.005}}
& \multirow{2}{*}{\makecell{$-$0.056\\/$-$0.039}}
& \multirow{2}{*}{$-$0.145}
& \multirow{2}{*}{\makecell{$-$0.201\\/$-$0.184}} \\ \cline{4-5}
&
&
& +20(47)$^{(a)}$
& \cite{deGroote2021}
&
&
&
&
&
& \\
\hline  
\ce{^{201}Hg}
& $\frac{3}{2}^-$
& $p_{3/2}$
& $-$0.346(35)
& \cite{mcdermott1960}
& \makecell{$-$1.148\\/$-$0.803}
& $-$0.332
& \makecell{$-$0.633\\/$-$0.446}
& \makecell{$-$0.329\\/$-$0.230}
& $-$0.101 & \makecell{$-$0.430\\/$-$0.332} \\
\hline

\ce{^{207}Po}
& $\frac{5}{2}^-$
& $f_{5/2}$
& +0.287(26)
& \cite{olsmats1961}
& \makecell{+0.547\\/+0.383}
& $-$0.338
& \makecell{+0.025\\/+0.005}
& \makecell{$-$0.433\\/$-$0.303}
& $-$0.236
& \makecell{$-$0.669\\/$-$0.539} \\
\hline

\end{tabular}
\label{tab:valence_neutron}
\end{table}

\clearpage


\scriptsize
\begin{table}[ht]
\caption{
Comparison between experimental ($exp$) and theoretical (single-particle ($s.p.$) and
collective strong ($scc$) and weak ($wcc$) coupling) values of magnetic octupole moments
for odd-proton nuclei.}
\centering
\begin{tabular}{|l|c|c|c|c|c|c|c|c|c|c|}
\hline
  \multicolumn{1}{|c|}{\multirow{3}{*}{Isotope}}
& \multicolumn{1}{c|} {\multirow{3}{*}{$I^\pi$}}
& \multicolumn{1}{c|} {\multirow{3}{*}{\makecell{Shell\\assign.}}}
& \multicolumn{1}{c|} {\multirow{2}{*}{Experimental}}
& \multicolumn{1}{c|} {\multirow{3}{*}{Ref.}}
& \multicolumn{1}{c|} {\multirow{2}{*}{\makecell{s.p.\\model}}}
& \multicolumn{5}{c|} {Collective model} \\ \cline{7-11}

&
&
&
&
&
& \multicolumn{2}{c|}{{\em Strong coupling}} & \multicolumn{3}{c|}{{\em Weak coupling}} \\ \cline{4-4}\cline{6-11}

&
&
& \multicolumn{1}{c|}{~$\Omega_{exp}/\langle r^2\rangle$} & & \multicolumn{1}{c|}{~$\Omega_{sp}/\langle r^2\rangle$}
& \multicolumn{1}{c|}{~$\Omega_{o}/\langle r^2\rangle$} & \multicolumn{1}{c|}{~$\Omega_{scc}/\langle r^2\rangle$}
& \multicolumn{1}{c|}{~$\Omega_{p}/\langle r^2\rangle$} & \multicolumn{1}{c|}{$\Omega_{c}/\langle r^2\rangle$}
& \multicolumn{1}{c|}{~$\Omega_{wcc}/\langle r^2\rangle$}\\
\hline

    
  \multicolumn{5}{|r|}{Equations used:~~}
& \multicolumn{1}{c|}{(\ref{eq:Omega shell model})}
& \multicolumn{1}{c|}{(\ref{eq:Omega0 for scc})}
& \multicolumn{1}{c|}{(\ref{eq:Omega value strong cc}) (\ref{eq:Omega value strong cc for I=3/2})}
& \multicolumn{1}{c|}{(\ref{eq:Omega_p value weak cc simplified version})}
& \multicolumn{1}{c|}{(\ref{eq:Omega_c value weak cc simplified version})}
& \multicolumn{1}{c|}{(\ref{eq:Omega break2})}\\
\hline
%
%
%
  \multirow{4}{*}{\ce{^{35}Cl}}
& \multirow{4}{*}{$\frac{3}{2}^+$}
& \multirow{4}{*}{$d_{3/2}$}
& $-0.1607(26)^{(a)}$
& \cite{schwartz1957}
& \multirow{4}{*}{\makecell{$-0.068$\\/$+0.021$}}
& \multirow{4}{*}{$-0.405$}
& \multirow{4}{*}{\makecell{$-0.048$\\/$-0.000$}}
& \multirow{4}{*}{\makecell{$-0.051$\\/$+0.016$}}
& \multirow{4}{*}{$-0.043$}
& ~~\multirow{4}{*}{\makecell{$-0.094$\\/$-0.027$}}~~ \\ \cline{4-4}
&
&
& $-$0.1607$^{(b)}$
& \cite{amoruso1971}
&
&
&
&
&
& \\ \cline{4-4}
&
&
& $-0.1385^{(b)}$
& \cite{amoruso1971}
&
&
&
&
&
& \\ \cline{4-4}
&
&
& $-0.1632(27)$
& \cite{zacharias1955}
&
&
&
&
&
& \\ \hline

  \multirow{2}{*}{\ce{^{37}Cl}}
& \multirow{2}{*}{$\frac{3}{2}^+$}
& \multirow{2}{*}{$d_{3/2}$}
& $-0.1202(25)^{(a)}$
& \cite{schwartz1957}
& \multirow{2}{*}{\makecell{$-0.068$\\/$+0.021$}}
& \multirow{2}{*}{$-0.383$}
& \multirow{2}{*}{\makecell{$-0.048$\\/$+0.000$}}
& \multirow{2}{*}{\makecell{$-0.051$\\/$+0.016$}}
& \multirow{2}{*}{$-0.040$}
& \multirow{2}{*}{\makecell{$-0.091$\\/$-0.024$}} \\ \cline{4-4}
&
&
& $-0.122(26)$
& \cite{zacharias1955}
&
&
&
&
&
& \\ \hline

\ce{^{45}Sc}
& $\frac{7}{2}^-$
& $f_{7/2}$
& $+12(6)$
& \cite{deGroote2022}
& \makecell{$+3.794$\\/$+3.055$}
& $-0.389$
& \makecell{$+0.722$\\/$+0.566$}
& \makecell{$+2.430$\\/$+1.957$}
& $-0.084$
& \makecell{$+2.347$\\/$+1.874$}\\ \hline

\multirow{2}{*}{\ce{^{69}Ga}}
& \multirow{2}{*}{$\frac{3}{2}^-$}
& \multirow{2}{*}{$p_{3/2}$}
& $+0.745(27)^{(a)}$
& \cite{schwartz1957}
& \multirow{2}{*}{\makecell{$+1.676$\\/$+1.173$}}
& \multirow{2}{*}{$-0.374$}
& \multirow{2}{*}{\makecell{$+0.899$\\/$+0.626$}}
& \multirow{2}{*}{\makecell{$+1.207$\\/$+0.845$}}
& \multirow{2}{*}{$-0.045$}
& \multirow{2}{*}{\makecell{$+1.162$\\/$+0.800$}}\\ \cline{4-4}
&
&
& $+0.58(11)$
& \cite{daly1954}
& 
& 
& 
& 
& 
& \\ \hline

\multirow{2}{*}{\ce{^{71}Ga}}
& \multirow{2}{*}{$\frac{3}{2}^-$}
&\multirow{2}{*}{$p_{3/2}$}
& $+0.960(27)^{(a)}$
& \cite{schwartz1957}
& \multirow{2}{*}{\makecell{$+1.676$\\/$+1.173$}}
& \multirow{2}{*}{$-0.364$}
& \multirow{2}{*}{\makecell{$+0.899$\\/$+0.627$}}
& \multirow{2}{*}{\makecell{$+1.217$\\/$+0.852$}}
& \multirow{2}{*}{$-0.043$}
& \multirow{2}{*}{\makecell{$+1.174$\\/$+0.809$}}\\ \cline{4-4}
&
&
& $+0.78(11)$
& \cite{daly1954}
& 
& 
& 
& 
& 
& \\ \hline

\multirow{3}{*}{\ce{^{79}Br}}
& \multirow{3}{*}{$\frac{3}{2}^-$}
& \multirow{3}{*}{$p_{3/2}$}
& $+0.461^{(a),(b)}$
& \cite{amoruso1971}
& \multirow{3}{*}{\makecell{$+1.676$\\/$+1.173$}}
& \multirow{3}{*}{$-0.369$}
& \multirow{3}{*}{\makecell{$+0.899$\\/$+0.626$}}
& \multirow{3}{*}{\makecell{$+1.183$\\/$+0.828$}}
& \multirow{3}{*}{$-0.047$}
& \multirow{3}{*}{\makecell{$+1.137$\\/$+0.782$}}\\ \cline{4-4}
&
&
& $+0.611^{(b)}$
& \cite{amoruso1971}
& 
& 
&
& 
& 
& \\ \cline{4-4}
&
&
& $+0.576^{(b)}$
& \cite{brown1966} 
& 
& 
& 
&
&
& \\ \hline

\ce{^{81}Br}
& $\frac{3}{2}^-$
& $p_{3/2}$ 
& $+0.630^{(b)}$
& \cite{brown1966}         
& \makecell{$+1.676$\\/$+1.173$}     
& $-0.360$   
& \makecell{$+0.900$\\/$+0.627$} 
& \makecell{$+1.194$\\/$+0.836$}   
& $-0.044$
& \makecell{$+1.149$\\/$+0.791$}\\ \hline

\ce{^{87}Rb}
& $\frac{3}{2}^-$
& $p_{3/2}$  
& $-2.7(18)$
& \cite{gerginov2009}           
& \makecell{$+1.676$\\/$+1.173$}     
& $-0.354$  
& \makecell{$+0.900$\\/$+0.627$} 
& \makecell{$+1.186$\\/$+0.830$}   
& $-0.044$ 
& \makecell{$+1.142$\\/$+0.786$}\\ \hline

\ce{^{113}In}     
& $\frac{9}{2}^+$
& $g_{9/2}$  
& $+2.25(6)$ 
& \cite{eck1957}        
& \makecell{$+4.684$\\/$+3.933$}     
& $-0.361$   
& \makecell{$+1.270$\\/$+1.049$} 
& \makecell{$+2.971$\\/$+2.495$}   
& $-0.111$ 
& \makecell{$+2.861$\\/$+2.385$}\\ \hline

\multirow{2}{*}{\ce{^{115}In}}     
& \multirow{2}{*}{$\frac{9}{2}^+$}
& \multirow{2}{*}{$g_{9/2}$}  
& $+1.84(4)^{(a)}$
& \cite{schwartz1957}    
& \multirow{2}{*}{\makecell{$+4.684$\\/$+3.933$}}     
& \multirow{2}{*}{$-0.355$}   
& \multirow{2}{*}{\makecell{$+1.271$\\/$+1.051$}} 
& \multirow{2}{*}{\makecell{$+3.011$\\/$+2.528$}}   
& \multirow{2}{*}{$-0.106$} 
& \multirow{2}{*}{\makecell{$+2.905$\\/$+2.422$}}\\ \cline{4-4}
& 
&
& $+2.18(5)$
& \cite{eck1957} 
& 
& 
& 
& 
& 
& \\ \hline

\multirow{4}{*}{\ce{^{127}I}}      
& \multirow{4}{*}{$\frac{5}{2}^+$}
& \multirow{4}{*}{$d_{5/2}$}  
& $+0.66(17)^{(a)}$
& \cite{schwartz1957}   
& \multirow{4}{*}{\makecell{$+2.823$\\/$+2.148$}}     
& \multirow{4}{*}{$-0.348$}   
& \multirow{4}{*}{\makecell{$+0.295$\\/$+0.214$}} 
& \multirow{4}{*}{\makecell{$+0.646$\\/$+0.492$}}   
& \multirow{4}{*}{$-0.104$} 
& \multirow{4}{*}{\makecell{$+0.542$\\/$+0.387$}}\\ \cline{4-4}
&
&
& $+1^{(b)}$
& \cite{jaccarino1954}
&
&
&
&
&
& \\ \cline{4-4}       
&
&
& $+0.959^{(b)}$
& \cite{amoruso1971}
&
& 
& 
& 
& 
& \\ \cline{4-4}    
&
& 
& $+0.604^{(b)}$
& \cite{amoruso1971}
& 
& 
& 
&
&
& \\ \hline

\ce{^{133}Cs}
& $\frac{7}{2}^+$
& $g_{7/2}$
& $+2.9(4)$
& \cite{gerginov2003}
& \makecell{$+0.094$\\/$+0.611$}     
& $-0.345$   
& \makecell{$-0.053$\\/$+0.057$}       
& \makecell{$+0.042$\\/$+0.271$}   
& $-0.115$
& \makecell{$-0.073$\\/$+0.156$} \\ \hline

\multirow{3}{*}{\ce{^{197}Au}}
& \multirow{3}{*}{$\frac{3}{2}^+$}
& \multirow{3}{*}{$d_{3/2}$} 
& $+0.0265(19)^{(a)}$
& \cite{blachman1967} 
& \multirow{3}{*}{\makecell{$-0.068$\\/$+0.021$}}            
& \multirow{3}{*}{$-$0.334}   
& \multirow{3}{*}{\makecell{$-0.046$\\/$+0.002$}}       
& \multirow{3}{*}{\makecell{$-0.020$\\/$+0.006$}}           
& \multirow{3}{*}{$-0.101$}
& \multirow{3}{*}{\makecell{$-0.121$\\/$-0.095$}}\\ \cline{4-4}
& 
&
& $+0.35^{(b)}$ 
& \cite{blachman1967} 
& 
& 
& 
& 
& 
& \\ \cline{4-4}
& 
&
& $+0.16(16)$
& \cite{blachman1967} 
& 
& 
& 
& 
& 
& \\ \hline

\multirow{3}{*}{\ce{^{209}Bi}}
& \multirow{3}{*}{$\frac{9}{2}^-$}
& \multirow{3}{*}{$h_{9/2}$}
& $+1.43(8)^{(a)}$
& \cite{landman1970}     
& \multirow{3}{*}{\makecell{$+0.415$\\/$+1.113$}}     
& \multirow{3}{*}{$-0.331$}   
& \multirow{3}{*}{\makecell{$+0.025$\\/$+0.230$}} 
& \multirow{3}{*}{\makecell{$+0.092$\\/$+0.246$}}   
& \multirow{3}{*}{$-0.216$} 
& \multirow{3}{*}{\makecell{$-0.124$\\/$+0.031$}}\\ \cline{4-4}
&
& 
& $+1.12^{(b)}$
& \cite{hull1970}
& 
& 
& 
& 
& 
& \\ \cline{4-4}
& 
& 
& $+1.61(21)$
& \cite{rosen1972} 
& 
& 
& 
& 
& 
& \\ \hline

\end{tabular}
\label{tab:valence_proton}
\end{table}
\normalsize

\clearpage


\begin{longtable}{|c|c|c|c|c|c|c|}
    \caption{
    Collective model predictions (strong and weak coupling) for isotopes with a (bare) valence proton in subshell $2p_{3/2}$. All values are quoted in $\mu_N$ units. See explanation of tables for more details.
    }
    \\\hline
  \multicolumn{1}{|c|}{\multirow{3}{*}{Isotope}}
& \multicolumn{1}{c|}{\multirow{3}{*}{$I^\pi$}}
& \multicolumn{5}{c|}{Collective Model}\\\cline{3-7}
&
& \multicolumn{2}{c|}{\em Strong coupling}
& \multicolumn{3}{c|}{\em Weak coupling}\\\cline{3-4}\cline{5-7}
&
& \multicolumn{1}{c|}{$\frac{\Omega_{o}}{\langle r^2\rangle}$}
& \multicolumn{1}{c|}{$\frac{\Omega_{scc}}{\langle r^2\rangle}$}
& \multicolumn{1}{c|}{$\frac{\Omega_{p}}{\langle r^2\rangle}$}
& \multicolumn{1}{c|}{$\frac{\Omega_{c}}{\langle r^2\rangle}$}
& \multicolumn{1}{c|}{$\frac{\Omega_{wcc}}{\langle r^2\rangle}$}
\\\hline
\endfirsthead

\hline \multicolumn{7}{|r|}{{Continued from previous page}} \\ \hline
\endhead

\hline \multicolumn{7}{|r|}{{Continued on next page}} \\ \hline
\endfoot

\hline
\endlastfoot

\ce{^{55}Cu}  & $\frac{3}{2}^-$ & -0.439 & +0.897 & +1.158 & -0.058 & +1.100 \\ \hline
\ce{^{57}Cu}  & $\frac{3}{2}^-$ & -0.424 & +0.898 & +1.173 & -0.055 & +1.119 \\ \hline
\ce{^{59}Cu}  & $\frac{3}{2}^-$ & -0.410 & +0.898 & +1.187 & -0.051 & +1.136 \\ \hline
\ce{^{61}Cu}  & $\frac{3}{2}^-$ & -0.396 & +0.899 & +1.200 & -0.048 & +1.151 \\ \hline
\ce{^{63}Cu}  & $\frac{3}{2}^-$ & -0.384 & +0.899 & +1.211 & -0.046 & +1.166 \\ \hline
\ce{^{65}Cu}  & $\frac{3}{2}^-$ & -0.372 & +0.899 & +1.222 & -0.043 & +1.178 \\ \hline
\ce{^{67}Cu}  & $\frac{3}{2}^-$ & -0.361 & +0.900 & +1.231 & -0.041 & +1.190 \\ \hline
\ce{^{69}Cu}  & $\frac{3}{2}^-$ & -0.350 & +0.900 & +1.240 & -0.039 & +1.201 \\ \hline
\ce{^{71}Cu}  & $\frac{3}{2}^-$ & -0.340 & +0.900 & +1.248 & -0.037 & +1.211 \\ \hline
\ce{^{73}Cu}  & $\frac{3}{2}^-$ & -0.331 & +0.900 & +1.256 & -0.036 & +1.220 \\ \hline
\ce{^{59}Ga}  & $\frac{3}{2}^-$ & -0.438 & +0.897 & +1.140 & -0.060 & +1.079 \\ \hline
\ce{^{61}Ga}  & $\frac{3}{2}^-$ & -0.423 & +0.898 & +1.156 & -0.056 & +1.099 \\ \hline
\ce{^{63}Ga}  & $\frac{3}{2}^-$ & -0.410 & +0.898 & +1.170 & -0.053 & +1.117 \\ \hline
\ce{^{65}Ga}  & $\frac{3}{2}^-$ & -0.397 & +0.899 & +1.183 & -0.050 & +1.133 \\ \hline
\ce{^{67}Ga}  & $\frac{3}{2}^-$ & -0.385 & +0.899 & +1.195 & -0.047 & +1.148 \\ \hline
\ce{^{73m}Ga} & $\frac{3}{2}^-$ & -0.354 & +0.900 & +1.226 & -0.041 & +1.185 \\ \hline
\ce{^{75}Ga}  & $\frac{3}{2}^-$ & -0.344 & +0.900 & +1.235 & -0.039 & +1.196 \\ \hline
\ce{^{77}Ga}  & $\frac{3}{2}^-$ & -0.335 & +0.900 & +1.243 & -0.037 & +1.206 \\ \hline
\ce{^{79}Ga}  & $\frac{3}{2}^-$ & -0.327 & +0.901 & +1.250 & -0.036 & +1.215 \\ \hline
\ce{^{61}As}  & $\frac{3}{2}^-$ & -0.451 & +0.897 & +1.101 & -0.066 & +1.034 \\ \hline
\ce{^{63}As}  & $\frac{3}{2}^-$ & -0.436 & +0.897 & +1.119 & -0.062 & +1.057 \\ \hline
\ce{^{65}As}  & $\frac{3}{2}^-$ & -0.423 & +0.898 & +1.136 & -0.058 & +1.078 \\ \hline
\ce{^{73}As}  & $\frac{3}{2}^-$ & -0.377 & +0.899 & +1.190 & -0.047 & +1.143 \\ \hline 
\ce{^{75}As}  & $\frac{3}{2}^-$ & -0.367 & +0.899 & +1.201 & -0.045 & +1.156 \\ \hline
    \ce{^{77}As}  & $\frac{3}{2}^-$ & -0.357 & +0.900 & +1.211 & -0.043 & +1.168 \\ \hline
    \ce{^{79}As}  & $\frac{3}{2}^-$ & -0.348 & +0.900 & +1.220 & -0.041 & +1.179 \\ \hline
    \ce{^{81}As}  & $\frac{3}{2}^-$ & -0.339 & +0.900 & +1.228 & -0.039 & +1.190 \\ \hline
    \ce{^{87}As}  & $\left(\frac{5}{2}^- \text{,} \frac{3}{2}^-\right)$ & -0.316 & +0.901 & +1.251 & -0.034 & +1.216 \\ \hline
    \ce{^{75}Br}  & $\frac{3}{2}^-$ & -0.389 & +0.899 & +1.159 & -0.051 & +1.108  \\ \hline
    \ce{^{77}Br}  & $\frac{3}{2}^-$ & -0.379 & +0.899 & +1.172 & -0.049 & +1.123  \\ \hline
    \ce{^{83}Br}  & $\frac{3}{2}^-$ & -0.351 & +0.900 & +1.204 & -0.042 & +1.161  \\ \hline
    \ce{^{85}Br}  & $\frac{3}{2}^-$ & -0.343 & +0.900 & +1.213 & -0.041 & +1.172  \\ \hline
    \ce{^{89}Br}  & $\left(\frac{3}{2}^- \text{,} \frac{5}{2}^-\right)$ & -0.328 & +0.901 & +1.229 & -0.037 & +1.192  \\ \hline
    \ce{^{73}Rb}  & $\frac{3}{2}^-$ & -0.422 & +0.898 & +1.092 & -0.063 & +1.029  \\ \hline 
    \ce{^{75}Rb}  & $\frac{3}{2}^-$ & -0.411 & +0.898 & +1.109 & -0.060 & +1.049  \\ \hline
    \ce{^{77}Rb}  & $\frac{3}{2}^-$ & -0.400 & +0.898 & +1.125 & -0.056 & +1.068  \\ \hline
    \ce{^{81}Rb}  & $\frac{3}{2}^-$ & -0.381 & +0.899 & +1.152 & -0.051 & +1.101  \\ \hline
    \ce{^{89}Rb}  & $\frac{3}{2}^-$ & -0.346 & +0.900 & +1.196 & -0.043 & +1.153  \\ \hline
    \ce{^{91}Rb}  & $\frac{3}{2}^-$ & -0.339 & +0.900 & +1.205 & -0.041 & +1.164  \\ \hline
    \ce{^{97m}Rb} & $\left(\frac{1}{2} \text{,} \frac{3}{2}\right)^-$ & -0.318 & +0.901 & +1.229 & -0.036 & +1.192  \\ \hline
    \ce{^{83m}Y}  & $\left(\frac{3}{2}^-\right)$ & -0.391 & +0.899 & +1.116 & -0.056 & +1.060  \\ \hline
    \ce{^{109}Nb} & $\frac{3}{2}^-$ & -0.313 & +0.901 & +1.211 & -0.037 & +1.173  \\ \hline
    \ce{^{111}Nb} & $\frac{3}{2}^-$ & -0.308 & +0.901 & +1.218 & -0.036 & +1.182  \\ \hline 
    \ce{^{113}Nb} & $\frac{3}{2}^-$ & -0.302 & +0.901 & +1.225 & -0.035 & +1.190  \\ \hline
   \ce{^{115}Nb} & $\frac{3}{2}^-$  & -0.297 & +0.901 & +1.231 & -0.034 & +1.198  \\ \hline
    \ce{^{105}Tc} & $\left(\frac{3}{2}^-\right)$ & -0.341 & +0.900 & +1.157 & -0.045 & +1.111  \\ \hline
    \ce{^{107}Tc} & $\left(\frac{3}{2}^-\right)$ & -0.335 & +0.900 & +1.167 & -0.044 & +1.123

    \label{tab:p2p3_2}
\end{longtable}


\clearpage


\begin{longtable}{|c|c|c|c|c|c|c|}
    \caption{
    Collective model predictions (strong and weak coupling) for isotopes with a (bare) valence proton in subshell $2p_{3/2}$. All values are quoted in $\mu_N$ units. See explanation of tables for more details.
    }
    \\\hline
  \multicolumn{1}{|c|}{\multirow{3}{*}{Isotope}}
& \multicolumn{1}{c|}{\multirow{3}{*}{$I^\pi$}}
& \multicolumn{5}{c|}{Collective Model}\\\cline{3-7}
&
& \multicolumn{2}{c|}{\em Strong coupling}
& \multicolumn{3}{c|}{\em Weak coupling}\\\cline{3-4}\cline{5-7}
&
& \multicolumn{1}{c|}{$\frac{\Omega_{o}}{\langle r^2\rangle}$}
& \multicolumn{1}{c|}{$\frac{\Omega_{scc}}{\langle r^2\rangle}$}
& \multicolumn{1}{c|}{$\frac{\Omega_{p}}{\langle r^2\rangle}$}
& \multicolumn{1}{c|}{$\frac{\Omega_{c}}{\langle r^2\rangle}$}
& \multicolumn{1}{c|}{$\frac{\Omega_{wcc}}{\langle r^2\rangle}$}
\\\hline
\endfirsthead

\hline \multicolumn{7}{|r|}{{Continued from previous page}} \\ \hline
\endhead

\hline \multicolumn{7}{|r|}{{Continued on next page}} \\ \hline
\endfoot

\hline
\endlastfoot

\ce{^{73m}As} & $\frac{9}{2}^+$ & -0.377 & +1.265 & +3.343 & -0.090 & +3.252 \\ \hline
\ce{^{75m}As} & $\frac{9}{2}^+$ & -0.367 & +1.268 & +3.372 & -0.086 & +3.286 \\ \hline
\ce{^{77m}As} & $\frac{9}{2}^+$ & -0.357 & +1.271 & +3.400 & -0.082 & +3.318 \\ \hline
\ce{^{79m}As} & $\left(\frac{9}{2}\right)^+$ & -0.348 & +1.273 & +3.425 & -0.078 & +3.347 \\ \hline
\ce{^{77m}Br} & $\frac{9}{2}^+$ & -0.379 & +1.264 & +3.293 & -0.094 & +3.199 \\ \hline
\ce{^{79m}Br} & $\frac{9}{2}^+$ & -0.369 & +1.267 & +3.324 & -0.090 & +3.235 \\ \hline
\ce{^{81m}Br} & $\frac{9}{2}^+$ & -0.360 & +1.270 & +3.353 & -0.086 & +3.268 \\ \hline
\ce{^{73m}Rb} & $\frac{9}{2}^+$ & -0.422 & +1.252 & +3.072 & -0.122 & +2.950 \\ \hline
\ce{^{81m}Rb} & $\frac{9}{2}^+$ & -0.381 & +1.264 & +3.238 & -0.098 & +3.140 \\ \hline
\ce{^{83m}Rb} & $\frac{9}{2}^+$ & -0.371 & +1.267 & +3.272 & -0.094 & +3.178 \\ \hline
\ce{^{85m}Rb} & $\frac{9}{2}^+$ & -0.363 & +1.269 & +3.303 & -0.090 & +3.213 \\ \hline
\ce{^{95m}Rb} & $\frac{9}{2}^+$ & -0.324 & +1.280 & +3.429 & -0.073 & +3.356 \\ \hline
\ce{^{83}Y}   & $\left(\frac{9}{2}^+\right)$ & -0.391 &	+1.261 & +3.140 & -0.108 & +3.032 \\ \hline
\ce{^{85m}Y}  & $\left(\frac{9}{2}\right)^+$ & -0.382 &	+1.263 & +3.179 & -0.103 & +3.076 \\ \hline
\ce{^{87m}Y}  & $\frac{9}{2}^+$ & -0.373 &	+1.266 & +3.214 & -0.098 & +3.116 \\ \hline
\ce{^{89m}Y}  & $\frac{9}{2}^+$ & -0.365 &	+1.268 & +3.248 & -0.094 & +3.154 \\ \hline
\ce{^{91m}Y}  & $\frac{9}{2}^+$ & -0.357 &	+1.271 & +3.279 & -0.090 & +3.189 \\ \hline
\ce{^{93m}Y}  & $\frac{9}{2}^+$ & -0.349 &	+1.273 & +3.307 & -0.086 & +3.221 \\ \hline
\ce{^{95m}Y}  & $\frac{9}{2}^+$ & -0.342 & +1.275 & +3.334 & -0.083 & +3.252 \\ \hline
\ce{^{97m}Y}  & $\frac{9}{2}^+$ & -0.335 &	+1.277 & +3.359 & -0.079 & +3.280 \\ \hline
\ce{^{79}Nb}  & $\frac{9}{2}^+$ & -0.432 & +1.249 & +2.863 & -0.141 & +2.722 \\ \hline
\ce{^{81}Nb}  & $\frac{9}{2}^+$ & -0.422 & +1.252 & +2.923 & -0.133 & +2.790 \\ \hline
\ce{^{83}Nb}  & $\frac{9}{2}^+$ & -0.412 & +1.255 & +2.977 & -0.126 & +2.851 \\ \hline
\ce{^{85}Nb}  & $\frac{9}{2}^+$ & -0.402 &	+1.258 & +3.026 & -0.119 & +2.907 \\ \hline
\ce{^{87m}Nb} & $\left(\frac{9}{2}\right)^+$ & -0.393 & +1.260 & +3.072 & -0.113 & +2.958 \\ \hline
\ce{^{89}Nb}  & $\left(\frac{9}{2}^+\right)$ & -0.384 & +1.263 & +3.113 & -0.108 & +3.005 \\ \hline
\ce{^{91}Nb}  & $\frac{9}{2}^+$ & -0.375 &	+1.265 & +3.152 & -0.103 & +3.049 \\ \hline
\ce{^{93}Nb}  & $\frac{9}{2}^+$ & -0.367 &	+1.268 & +3.188 & -0.098 & +3.089 \\ \hline
\ce{^{95}Nb}  & $\frac{9}{2}^+$ & -0.360 &	+1.270 & +3.221 & -0.094 & +3.127 \\ \hline
\ce{^{97}Nb}  & $\frac{9}{2}^+$ & -0.352 &	+1.272 & +3.252 & -0.090 & +3.161 \\ \hline
\ce{^{99}Nb}  & $\frac{9}{2}^+$ & -0.345 &	+1.274 & +3.280 & -0.087 & +3.194 \\ \hline
\ce{^{87}Tc}  & $\frac{9}{2}^+$ & -0.412 &	+1.255 & +2.895 & -0.132 & +2.763 \\ \hline
\ce{^{89}Tc}  & $\left(\frac{9}{2}^+\right)$ & -0.403 &	+1.257 & +2.948 & -0.125 & +2.823 \\ \hline
\ce{^{91}Tc}  & $\left(\frac{9}{2}\right)^+$ & -0.394 &	+1.260 & +2.997 & -0.119 & +2.878 \\ \hline
\ce{^{93}Tc}  & $\frac{9}{2}^+$ & -0.385 &	+1.263 & +3.042 & -0.113 & +2.929 \\ \hline
\ce{^{95}Tc}  & $\frac{9}{2}^+$ & -0.377 &	+1.265 & +3.083 & -0.108 & +2.975 \\ \hline
\ce{^{97}Tc}  & $\frac{9}{2}^+$ & -0.369 &	+1.267 & +3.122 & -0.103 & +3.019 \\ \hline
\ce{^{99}Tc}  & $\frac{9}{2}^+$ & -0.362 &	+1.269 & +3.157 & -0.099 & +3.059 \\ \hline
\ce{^{101}Tc} & $\frac{9}{2}^+$ & -0.355 &	+1.271 & +3.191 & -0.095 & +3.096 \\ \hline
\ce{^{89}Rh}  & $\frac{9}{2}^+$ & -0.421 & +1.252 & +2.741 & -0.146 & +2.595 \\ \hline
\ce{^{91}Rh}  & $\left(\frac{9}{2}^+\right)$ & -0.412 & +1.255 & +2.804 & -0.138 & +2.666 \\ \hline
\ce{^{93}Rh}  & $\frac{9}{2}^+$ & -0.403 &	+1.257 & +2.862 & -0.131 & +2.730 \\ \hline
\ce{^{95}Rh}  & $\left(\frac{9}{2}\right)^+$ & -0.395 & +1.260 & +2.915 & -0.125 & +2.790 \\ \hline
\ce{^{97}Rh}  & $\frac{9}{2}^+$ & -0.387 & +1.262 & +2.963 & -0.119 & +2.844 \\ \hline
\ce{^{99m}Rh} & $\frac{9}{2}^+$ & -0.379 & +1.264 & +3.008 & -0.113 & +2.895 \\ \hline
\ce{^{101m}}Rh& $\frac{9}{2}^+$ & -0.371 & +1.267 & +3.050 & -0.108 & +2.941 \\ \hline
\ce{^{93}Ag}  & $\frac{9}{2}^+$ & -0.421 & +1.252 & +2.635 & -0.154 & +2.481 \\ \hline
\ce{^{95}Ag}  & $\left(\frac{9}{2}^+\right)$ & -0.412 & +1.255 & +2.704 & -0.146 & +2.558 \\ \hline
\ce{^{97}Ag}  & $\left(\frac{9}{2}\right)^+$ & -0.404 & +1.257 & +2.766 & -0.138 & +2.628 \\ \hline
\ce{^{99}Ag}  & $\left(\frac{9}{2}\right)^+$ & -0.396 & +1.259 & +2.824 & -0.132 & +2.692 \\ \hline
\ce{^{101}Ag} & $\frac{9}{2}^+$ & -0.388 & +1.262 & +2.877 & -0.125 & +2.751 \\ \hline
\ce{^{125}Ag} & $\left(\frac{9}{2}^+\right)$ & -0.313 & +1.284 & +3.288 & -0.078 & +3.209 \\ \hline
\ce{^{127}Ag} & $\left(\frac{9}{2}^+\right)$ & -0.308 & +1.285 & +3.310 & -0.076 & +3.234 \\ \hline
\ce{^{129}Ag} & $\frac{9}{2}^+$ & -0.304 & +1.287 & +3.332 & -0.073 & +3.258 \\ \hline
\ce{^{131}Ag} & $\frac{9}{2}^+$ & -0.299 & +1.288 & +3.352 & -0.071 & +3.281 \\ \hline
\ce{^{133}Ag} & $\frac{9}{2}^+$ & -0.294 & +1.289 & +3.371 & -0.069 & +3.302 \\ \hline
\ce{^{97}In}  & $\frac{9}{2}^+$ & -0.421 & +1.252 & +2.518 & -0.163 & +2.355 \\ \hline
\ce{^{99}In}  & $\frac{9}{2}^+$ & -0.412 & +1.255 & +2.593 & -0.154 & +2.438 \\ \hline
\ce{^{101}In} & $\left(\frac{9}{2}^+\right)$ & -0.404 & +1.257 & +2.661 & -0.146 & +2.515 \\ \hline
\ce{^{103}In} & $\left(\frac{9}{2}^+\right)$ & -0.396 & +1.259 & +2.724 & -0.139 & +2.585 \\ \hline
\ce{^{105}In} & $\frac{9}{2}^+$ & -0.389 & +1.261 & +2.781 & -0.132 & +2.649 \\ \hline
\ce{^{107}In} & $\frac{9}{2}^+$ & -0.382 & +1.264 & +2.834 & -0.126 & +2.708 \\ \hline
\ce{^{109}In} & $\frac{9}{2}^+$ & -0.375 & +1.266 & +2.883 & -0.121 & +2.763 \\ \hline
\ce{^{111}In} & $\frac{9}{2}^+$ & -0.368 & +1.268 & +2.929 & -0.115 & +2.814 \\ \hline
\ce{^{117}In} & $\frac{9}{2}^+$ & -0.349 & +1.273 & +3.048 & -0.102 & +2.946 \\ \hline
\ce{^{119}In} & $\frac{9}{2}^+$ & -0.343 & +1.275 & +3.083 & -0.098 & +2.985 \\ \hline
\ce{^{121}In} & $\frac{9}{2}^+$ & -0.337 & +1.277 & +3.115 & -0.095 & +3.021 \\ \hline
\ce{^{123}In} & $\frac{9}{2}^+$ & -0.332 & +1.278 & +3.146 & -0.091 & +3.055 \\ \hline
\ce{^{125}In} & $\frac{9}{2}^+$ & -0.327 & +1.280 & +3.175 & -0.088 & +3.087 \\ \hline
\ce{^{127}In} & $\frac{9}{2}^+$ & -0.321 & +1.281 & +3.202 & -0.085 & +3.117 \\ \hline
\ce{^{129}In} & $\frac{9}{2}^+$ & -0.316 & +1.283 & +3.228 & -0.082 & +3.146 \\ \hline
\ce{^{131}In} & $\frac{9}{2}^+$ & -0.312 & +1.284 & +3.252 & -0.080 & +3.173 \\ \hline
\ce{^{133}In} & $\left(\frac{9}{2}^+\right)$ & -0.307 & +1.286 & +3.276 & -0.077 & +3.198 \\ \hline
\ce{^{135}In} & $\frac{9}{2}^+$ & -0.302 & +1.287 & +3.298 & -0.075 & +3.223 \\ \hline
\ce{^{137}In} & $\frac{9}{2}^+$ & -0.298 & +1.288 & +3.319 & -0.073 & +3.246

\label{tab:p1g9_2}
\end{longtable}


\clearpage


\begin{longtable}{|c|c|c|c|c|c|c|}
    \caption{
    Collective model predictions (strong and weak coupling) for isotopes with a (bare) valence neutron in subshell $1g_{9/2}$. All values are quoted in $\mu_N$ units. See explanation of tables for more details.
    }
    \\\hline
  \multicolumn{1}{|c|}{\multirow{3}{*}{Isotope}}
& \multicolumn{1}{c|}{\multirow{3}{*}{$I^\pi$}}
& \multicolumn{5}{c|}{Collective Model}\\\cline{3-7}
&
& \multicolumn{2}{c|}{\em Strong coupling}
& \multicolumn{3}{c|}{\em Weak coupling}\\\cline{3-4}\cline{5-7}
&
& \multicolumn{1}{c|}{$\frac{\Omega_{o}}{\langle r^2\rangle}$}
& \multicolumn{1}{c|}{$\frac{\Omega_{scc}}{\langle r^2\rangle}$}
& \multicolumn{1}{c|}{$\frac{\Omega_{p}}{\langle r^2\rangle}$}
& \multicolumn{1}{c|}{$\frac{\Omega_{c}}{\langle r^2\rangle}$}
& \multicolumn{1}{c|}{$\frac{\Omega_{wcc}}{\langle r^2\rangle}$}
\\\hline
\endfirsthead

\hline \multicolumn{7}{|r|}{{Continued from previous page}} \\ \hline
\endhead

\hline \multicolumn{7}{|r|}{{Continued on next page}} \\ \hline
\endfoot

\hline
\endlastfoot

\ce{^{59m}Cr} & $\left(\frac{9}{2}^+\right)$ & -0.339 & -0.712 & -1.588 & -0.068 & -1.656 \\ \hline
\ce{^{61m}Fe} & $\frac{9}{2}^+$ & -0.355 & -0.717 & -1.564 & -0.074 & -1.639 \\ \hline
\ce{^{65m}Fe} & $\left(\frac{9}{2}^+\right)$ & -0.333 & -0.711 & -1.584 & -0.067 & -1.651 \\ \hline
\ce{^{75}Fe}  & $\frac{9}{2}^+$ & -0.289 & -0.698 & -1.622 & -0.054 & -1.675 \\ \hline
\ce{^{67m}Ni} & $\frac{9}{2}^+$ & -0.348 & -0.715 & -1.559 & -0.074 & -1.633 \\ \hline      
\ce{^{69}Ni}  & $\left(\frac{9}{2}^+\right)$ & -0.338 & -0.712 & -1.569 & -0.070 & -1.639 \\ \hline
\ce{^{71}Ni}  & $\left(\frac{9}{2}^+\right)$ & -0.329 & -0.709 & -1.578 & -0.067 & -1.645 \\ \hline
\ce{^{73}Ni}  & $\left(\frac{9}{2}^+\right)$ & -0.320 & -0.707 & -1.587 & -0.064 & -1.651 \\ \hline
\ce{^{75}Ni}  & $\frac{9}{2}^+$ & -0.311 & -0.704 & -1.595 & -0.061 & -1.656 \\ \hline
\ce{^{77}Ni}  & $\frac{9}{2}^+$ & -0.303 & -0.702 & -1.602 & -0.059 & -1.661 \\ \hline
\ce{^{69m}Zn} & $\frac{9}{2}^+$ & -0.362 & -0.719 & -1.531 & -0.081 & -1.612 \\ \hline
\ce{^{71m}Zn} & $\frac{9}{2}^+$ & -0.352 & -0.716 & -1.543 & -0.077 & -1.620 \\ \hline
\ce{^{79}Zn}  & $\frac{9}{2}^+$ & -0.316 & -0.706 & -1.580 & -0.064 & -1.644 \\ \hline
\ce{^{71m}Ge} & $\frac{9}{2}^+$ & -0.376 & -0.723 & -1.500 & -0.088 & -1.588 \\ \hline
\ce{^{73}Ge}  & $\frac{9}{2}^+$ & -0.365 & -0.720 & -1.513 & -0.084 & -1.597 \\ \hline
\ce{^{81}Ge}  & $\frac{9}{2}^+$ & -0.329 & -0.710 & -1.555 & -0.070 & -1.625 \\ \hline
\ce{^{73}Se}  & $\frac{9}{2}^+$ & -0.388 & -0.727 & -1.464 & -0.097 & -1.561  \\ \hline
\ce{^{83}Se}  & $\frac{9}{2}^+$ & -0.341 & -0.713 & -1.527 & -0.077 & -1.604 \\ \hline
\ce{^{73m}Kr} & $\left(\frac{9}{2}^+\right)$ & -0.411 & -0.734 & -1.404 & -0.113 & -1.516 \\ \hline 
\ce{^{85}Kr}  & $\frac{9}{2}^+$ & -0.353 & -0.717 & -1.496 & -0.084 & -1.579 \\ \hline 
\ce{^{85}Sr}  & $\frac{9}{2}^+$ & -0.372 & -0.722 & -1.445 & -0.096 & -1.541 \\ \hline 
\ce{^{87}Sr}  & $\frac{9}{2}^+$ & -0.364 & -0.720 & -1.460 & -0.092 & -1.551 \\ \hline
\ce{^{87}Zr}  & $\frac{9}{2}^+$ & -0.383 & -0.725 & -1.402 & -0.105 & -1.507 \\ \hline
\ce{^{89}Zr}  & $\frac{9}{2}^+$ & -0.374 & -0.723 & -1.419 & -0.100 & -1.519  \\ \hline
\ce{^{89}Mo}  & $\left(\frac{9}{2}^+\right)$ & -0.393 & -0.728 & -1.352 & -0.116 & -1.468 \\ \hline 
\ce{^{91}Mo}  & $\frac{9}{2}^+$ & -0.385 & -0.726 & -1.372 & -0.110 & -1.482  \\ \hline 
\ce{^{89}Ru}  & $\left(\frac{9}{2}^+\right)$ & -0.412 & -0.734 & -1.270 & -0.135 & -1.405 \\ \hline
\ce{^{91}Ru}  & $\left(\frac{9}{2}^+\right)$ & -0.403 & -0.731 & -1.295 & -0.128 & -1.423 \\ \hline
\ce{^{93}Ru}  & $\left(\frac{9}{2}\right)^+$ & -0.394 & -0.729 & -1.317 & -0.122 & -1.439  \\ \hline
\ce{^{93}Pd}  & $\left(\frac{9}{2}^+\right)$ & -0.412 & -0.734 & -1.228 & -0.142 & -1.370 \\ \hline
\ce{^{95}Pd}  & $\frac{9}{2}^+$ & -0.403 & -0.731 & -1.254 & -0.135 & -1.389 \\ \hline
\ce{^{95}Cd}  & $\frac{9}{2}^+$ & -0.421 & -0.737 & -1.149 & -0.158 & -1.307 \\ \hline
\ce{^{97}Cd}  & $\left(\frac{9}{2}^+\right)$ & -0.412 & -0.734 & -1.181 & -0.150 & -1.331 \\ \hline
\ce{^{99}Sn}  & $\frac{9}{2}^+$ & -0.421 & -0.737 & -1.093 & -0.168 & -1.261
    
    \label{tab:n1g9_2}
\end{longtable}

\normalsize

\end{document}